\documentclass[aps,prb,floatfix,twocolumn,notitlepage,superscriptaddress,10pt]{revtex4-2}

\usepackage{xcolor}
\usepackage{amsmath, amsthm, amssymb, bbold}
\usepackage[normalem]{ulem}
\usepackage{cancel}

\usepackage{graphicx}
\usepackage{bm}
\usepackage{microtype}
\usepackage{verbatim}   % useful for program listings
\usepackage{color}      % use if color is used in text
\usepackage{microtype}
\usepackage[normalem]{ulem}
\usepackage{enumitem} 
\usepackage{dsfont} 
\usepackage[T1]{fontenc}
\usepackage{hyperref}
\hypersetup{colorlinks,
	linkcolor=blue,
	urlcolor=blue,
	citecolor=blue
}

\newcommand{\bra}[1]{\langle #1|}
\newcommand{\ket}[1]{|#1\rangle}

\newcommand{\braket}[1]{\langle #1\rangle}

\begin{document}
\title{Nonstabilizerness as a diagnostic of criticality and exceptional points in non-Hermitian spin chains}
\author{C\u at\u alin Pa\c scu Moca}
%	\email{mocap@uoradea.ro}
\affiliation{Department of Physics, University of Oradea, RO-410087 Oradea, Romania}
\affiliation{Department of Theoretical Physics, Institute of Physics, Budapest University of Technology and Economics, M\H{u}egyetem rkp.~3, H-1111 Budapest, Hungary}
\affiliation{MTA-BME Lend\"ulet ``Momentum'' Open Quantum Systems Research Group, Institute of Physics, Budapest University of Technology and Economics,
	M\H uegyetem rkp.~3, H-1111, Budapest, Hungary}
% \orcid{https://orcid.org/0000-0002-1299-399X}
\author{Doru Sticlet}
\email{doru.sticlet@itim-cj.ro}
\affiliation{National Institute for R\&D of Isotopic and Molecular Technologies, RO-400293 Cluj-Napoca, Romania}
% \orcid{https://orcid.org/0000-0003-0646-1978}
\author{Bal\'azs D\'ora}
\affiliation{Department of Theoretical Physics, Institute of Physics, Budapest University of Technology and Economics, M\H{u}egyetem rkp.~3, H-1111 Budapest, Hungary}
\affiliation{MTA-BME Lend\"ulet ``Momentum'' Open Quantum Systems Research Group, Institute of Physics, Budapest University of Technology and Economics,
	M\H uegyetem rkp.~3, H-1111, Budapest, Hungary}
% \orcid{https://orcid.org/0000-0002-5984-3761}
%	\date{\today}

\begin{abstract}
	We investigate nonstabilizerness, also known as ``magic,'' to understand criticality and exceptional points in non-Hermitian quantum many-body systems. Our focus is on parity-time ($\mathcal{PT}$)-symmetric spin chains, specifically the non-Hermitian transverse-field Ising and XX models. We calculate stabilizer Rényi entropies in their ground states using non-Hermitian matrix product state methods. Our findings show that magic exhibits unique and model-specific signs of phase transitions. In the Ising chain, it peaks near the regular Hermitian-like critical line but disappears across exceptional points. 
	In contrast, in the XX chain, it reaches its maximum at the exceptional line where $\mathcal{PT}$ symmetry is broken. 
	Finite-size scaling reveals that these effects become more pronounced with larger systems, highlighting nonstabilizerness as a sensitive marker for both quantum criticality and non-Hermitian spectral degeneracies.
	We also investigate magic in momentum space for the XX model analytically and find that it reaches a minimum around exceptional points. 
	Our results indicate that magic takes extremal values at the exceptional points and serves as a valuable tool for examining complexity,
	criticality, and symmetry breaking in non-Hermitian quantum matter.
\end{abstract}

\maketitle

\section{Introduction}

Non-Hermitian quantum systems have recently become a central theme in condensed matter
physics, statistical mechanics, and quantum optics~\cite{moiseyev2011non,ashida2020non,wang2023non}.
While Hermitian Hamiltonians guarantee
real spectra and unitary dynamics, non-Hermitian Hamiltonians relax these constraints,
allowing for complex energy spectra and nonunitary evolution~\cite{rotter2009non,roccati2022non,gomez2022bridging}.
Such models naturally arise in engineered photonic~\cite{yan2023advances,li2023exceptional} and cold-atom platforms~\cite{li2020topological,luo2024quantum}.
A particular subclass is formed by $\mathcal{PT}$ symmetric Hamiltonians, which have real
spectra in the $\mathcal{PT}$-symmetric regime, while crossing an exceptional point drives them into the $\mathcal{PT}$-broken phase, where eigenvalues appear in complex-conjugate pairs~\cite{mostafazadeh2002pseudo,mostafazadeh2003exact,feng2017non,longhi2018parity,Bender2024}.

In parallel, quantum resource theory~\cite{chitambar2019quantum} has introduced new perspectives on many-body states,
going beyond conventional order parameters and entanglement
entropy~\cite{veitch2014resource,howard2017application,liu2022many}.
One such concept is nonstabilizerness, often referred to as ``magic''~\cite{leone2022stabilizer,odavic2024stabilizer,tarabunga2024critical,haug2023stabilizer,sinibaldi2025non,huang2024non}.
Magic quantifies the distance of a quantum
state from the set of stabilizer states, which form the cornerstone of classically simulable
quantum computation and stabilizer error correction codes~\cite{gottesman1997stabilizer,Aaronson2004,dehaene2003clifford,masot2024stabilizer}.
A nonzero amount of magic is indispensable for universal fault-tolerant quantum computation, as it enables the realization of non-Clifford gates~\cite{brown2020fault}.
In many-body settings, magic has emerged as a sensitive probe of correlations, scrambling, and criticality, complementing entanglement measures and
revealing new aspects of quantum complexity~\cite{oliviero2022magic,odavic2023complexity,
sticlet2025non,sarkis2025molecules,moca2025non,tarabunga2024critical,Iannotti2025,Turkeshi2025a,bera2025non,Viscardi2025,Catalano2025}.

Despite rapid progress in both directions, the intersection of
non-Hermitian physics and nonstabilizerness remains largely
unexplored.
Non-Hermitian Hamiltonians are characterized by non-orthogonal
eigenstates and exceptional points~\cite{heiss2004exceptional,Berry2004,minganti2019quantum, ding2022non}, raising fundamental questions about
how quantum information measures behave in such contexts.
For instance, does magic detect the onset of $\mathcal{PT}$-symmetry
breaking and can nonstabilizerness provide a universal diagnostic of
non-Hermitian quantum criticality or exceptional points?

In this work, we address these questions by investigating the
stabilizer R\'enyi entropies in the ground state for two setups: for the non-Hermitian
transverse-field Ising chain~\cite{starkov2023quantum, lu2024many, Tirrito2025} and
for the non-Hermitian XX model~\cite{Ashida2017,dora2022correlations}.
We demonstrate that nonstabilizerness offers a powerful tool to identify both conventional quantum critical points and non-Hermitian exceptional points.
Our analysis shows that magic takes extremal values at the exceptional points, both in real space and in momentum space.
These results position nonstabilizerness as a valuable addition to the growing toolbox of probes for non-Hermitian quantum matter.

Beyond these motivations, studying nonstabilizerness in non-Hermitian settings may also provide insights into the resource-based aspects of quantum computation under nonunitary evolution~\cite{fortin2016non,williams2004probabilistic,turkeshi2023entanglement}.
As experimental realizations of non-Hermitian systems become more common in various setups, understanding how magic works in these environments is directly relevant for near-term platforms.

\section{Stabilizer R\'enyi entropy }

To quantify nonstabilizerness (``magic'') in the ground state, we work in the right-right (natural self-normal) basis using open boundary conditions (OBC).
Notably, alternative approaches in non-Hermitian physics employ a left-right biorthogonal basis~\cite{ashida2020non}. 
While mathematically elegant, the biorthogonal approach is known to lead to difficulties in the physical interpretation of quantities such as expectation values and correlation functions~\cite{Meden2023}. 
The right-right basis, on the other hand, is the standard approach for effective non-Hermitian Hamiltonians describing open quantum systems modeled by Lindblad master equations in the absence of quantum jumps or in the presence of postselection~\cite{Dalibard1992,Carmichael1993}.
The ground state of the non-Hermitian Hamiltonian is computed using a density matrix renormalization group (DMRG) algorithm~\cite{White1992} adapted for non-Hermitian systems, as implemented in the \texttt{ITensor} library~\cite{fishman2022itensor}.
The right eigenstates are represented as matrix product states (MPS), with open boundary conditions~\cite{schollwock2011density}.
Standard DMRG sweeps are modified to account for the non-Hermitian character
of $H$, ensuring convergence to the correct right eigenvector.
The ground state is identified as the state minimizing the real part of the energy and convergence is checked by monitoring the absolute value of the ground-state energy.
In simulations a maximum bond dimension $\chi=256$ is used.

% Denoting the right ground state of the Hamiltonian $H$ as $\ket{\psi_R}$, we define, for a subregion $A$ of length $\ell$, and its complement $\bar A$, the reduced density matrix
% \begin{equation}
% 	\rho^{RR}_A = \frac{\mathrm{Tr}_{\bar A} \left( \ket{\psi_R}\bra{\psi_R} \right)}{\braket{\psi_R|\psi_R}},
% \end{equation}
% which depends solely on the right eigenvector, and coincides with the standard Hermitian reduced density matrix when $H$ is Hermitian.
% Expanding $\rho^{RR}_A$ in the $\ell$-qubit Pauli string basis $\mathcal{P}_\ell$,
% \begin{equation}
% 	\rho^{RR}_A = \frac{1}{2^\ell} \sum_{P \in \mathcal{P}_\ell} c_P  P, \qquad c_P = \mathrm{Tr}(\rho^{RR}_A P),
% \end{equation}
% the stabilizer $2$-Rényi magic in the right–right basis is defined as
% \begin{equation}\label{eq:M2RR}
% 	M_2^{RR}(A) = -\log_2 \left( \sum_{P \in \mathcal{P}_\ell} |c_P|^4 \right).
% \end{equation}
% In practice, the coefficients $c_P$ are efficiently obtained by evaluating expectation values of Pauli strings in the right-right basis.

Denoting the right ground state of the Hamiltonian $H$ as $\ket{\psi_R}$, we define the system density matrix
\begin{equation}
	\rho^{RR} =  \frac{\ket{\psi_R}\bra{\psi_R}}{\braket{\psi_R|\psi_R}},
\end{equation}
which depends solely on the right eigenvector, and coincides with the standard Hermitian reduced density matrix when $H$ is Hermitian.
Expanding $\rho^{RR}$ in the $L$-qubit Pauli string basis $\mathcal{P}_L$,
\begin{equation}
	\rho^{RR} = \frac{1}{2^L} \sum_{P \in \mathcal{P}_L} c_P  P, \qquad c_P = \mathrm{Tr}(\rho^{RR} P),
\end{equation}
the stabilizer $2$-Rényi magic in the right–right basis is defined as
\begin{equation}\label{eq:M2RR}
	M_2^{RR} = -\log_2 \left( \sum_{P \in \mathcal{P}_L} |c_P|^4 \right).
\end{equation}
In practice, the coefficients $c_P$ are efficiently obtained by evaluating expectation values of Pauli strings in the right-right basis.

To  compute the stabilizer $2$-Rényi magic $M_2^{RR}$, we employ a Metropolis-Hastings sampling scheme~\cite{chib1995understanding,robert2009metropolis,roberts2001optimal} within the MPS framework.
The idea is to sample Pauli strings $P \in \mathcal{P}_L$ according to the weight $|c_P|^2$.
Starting from an initial Pauli string, candidate updates are generated by single-qubit flips or local Pauli substitutions, and accepted with
probability proportional to the ratio of $|c_P|^2$ for the new and old strings, ensuring convergence to the correct distribution~\cite{Tarabunga2023}.
In the MPS language~\cite{haug2023stabilizer}, each evaluation of $c_P$ is performed by contracting the MPS
tensors in the RR basis with the corresponding Pauli operator on subsystem $A$.
This approach allows for scalable estimation of $M_2^{RR}$ in large $L$ limit, where enumerating all $4^L$ Pauli strings would be
computationally prohibitive.
In our simulations, the number of samples for the Metropolis-Hastings algorithm is typically on the order of $10^5$.
To minimize the errors, we perform an average over several separate Markov chains.

We stress that stabilizer R\'enyi entropies provide information complementary to entanglement entropies.
While entanglement captures correlations across bipartitions, magic measures the intrinsic non-classicality of the state in terms of its departure from stabilizer form.
This is equally true in Hermitian and non-Hermitian systems.
Specific to the non-Hermitian systems is the presence of exceptional points where eigenstates coalesce.
Entanglement can survive at these points, but magic can vanish if the state approaches a generalized stabilizer structure.
%In non-Hermitian systems, this distinction becomes crucial: entanglement may survive even when eigenstates acquire non-orthogonal character, but magic can vanish if the state approaches a generalized stabilizer structure.
Thus, analyzing $M_2^{RR}$ reveals unique signatures of non-Hermitian quantum matter.

\begin{figure}[t]
	\begin{center}
		\includegraphics[width=0.95\columnwidth]{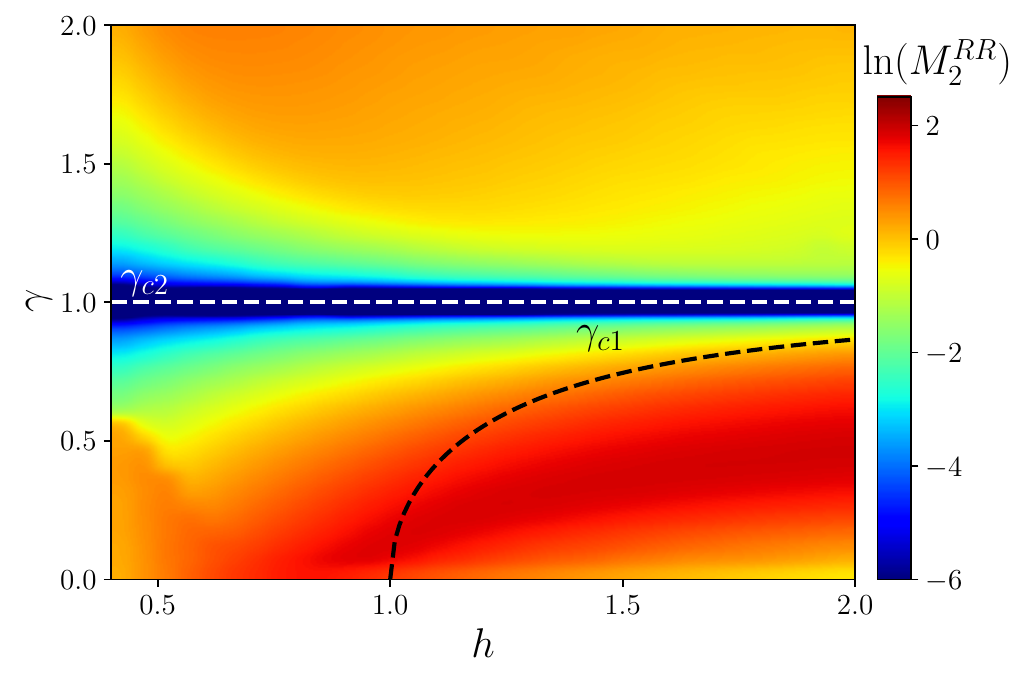}
		\caption{Ground-state density plot of $M_2^{RR}$. The dashed black and white lines indicate the
		theoretical critical lines $\gamma_{c1}$ and $\gamma_{c2}$ in the thermodynamic limit.
		Magic peaks along the Ising transition line and vanishes at the exceptional points across the phase diagram. System size is fixed to $L=16$.
		}
		\label{fig:phase_diagram}
	\end{center}
\end{figure}

\section{Non-hermitian Transverse field Ising model}

The one-dimensional non-Hermitian transverse-field Ising (NHTI) model~\cite{lu2024many}, defined by
\begin{equation}
	H = -J \sum_{j=1}^{L}  \sigma_j^x \sigma_{j+1}^x + \sum_{j=1}^{L} h(\sigma_j^z + i\gamma \sigma_j^y),
	\label{eq:H}
\end{equation}
extends the conventional Hermitian transverse-field Ising chain by introducing an imaginary
field along the $y$ direction.
For $\gamma=0$ the model reduces to the standard one~\cite{pfeuty1970one}.
In this limit, it undergoes a second-order quantum phase transition between ferromagnetic (FM) and paramagnetic (PM) phase at $h=J$.
When $\gamma \neq 0$, the Hamiltonian becomes non-Hermitian but retains $\mathcal{PT}$ symmetry, ensuring that eigenvalues are either real or come in complex-conjugate pairs.
The model is exactly solvable via a Jordan-Wigner transformation, which maps it to a non-Hermitian free-fermion problem~\cite{lu2024many}.

For non-Hermitian Hamiltonians, the notion of a ground state is not uniquely defined due to the possibility of complex eigenvalues.
In this work, we define the ground state as the eigenstate with the smallest real part of the energy, providing a consistent convention for analyzing stabilizer R\'enyi entropy.

The ground state phase diagram in the $(h,\gamma)$ plane reveals two distinct critical lines.
The first critical line corresponds to the \emph{ground-state Ising transition} in the $\mathcal{PT}$ symmetric region~\cite{yang2022hidden} for $\gamma<1$, occurring at
\begin{equation}
	\gamma_{c1} = \sqrt{1 - (J/h)^2}.
\end{equation}
The second critical is the \emph{full $\mathcal{PT}
	$ transition} at $\gamma_{c2} = 1$, where the entire many-body spectrum changes
from $\mathcal{PT}$-symmetric (real energies) to $\mathcal{PT}$-broken (complex energies),
featuring second-order exceptional points \footnote{An analysis of the excited states
	reveals the presence of the third critical line,  in the $\mathcal{PT}$-broken regime
	($\gamma>1$), where the first and second excited states coalesce and acquire complex
	energies, again through second-order exceptional points.}.
The two critical lines as emerging in the thermodynamic limit, are indicated by the dashed lines in Fig.~\ref{fig:phase_diagram}.
By symmetry, the phase diagram extends naturally to
negative values of both $h$ and $\gamma$, covering the regions $h<0$ and $\gamma<0$ as well.

The FM and PM phases are separated by these critical lines,
with $\mathcal{PT}$-symmetric regions hosting purely real energies and $\mathcal{PT}
$-broken regions exhibiting finite imaginary parts in the spectrum~\cite{mostafazadeh2002pseudo}.
The non-Hermitian character modifies the quasiparticle dispersion relation by introducing complex-valued energies, and this affects the dynamical correlations and observables. The interplay of the transverse field $h$ and the non-Hermitian parameter $\gamma$ provides a rich structure that we exploit to study nonstabilizerness.

\begin{figure}[th!]
	\begin{center}
		\includegraphics[width=0.95\columnwidth]{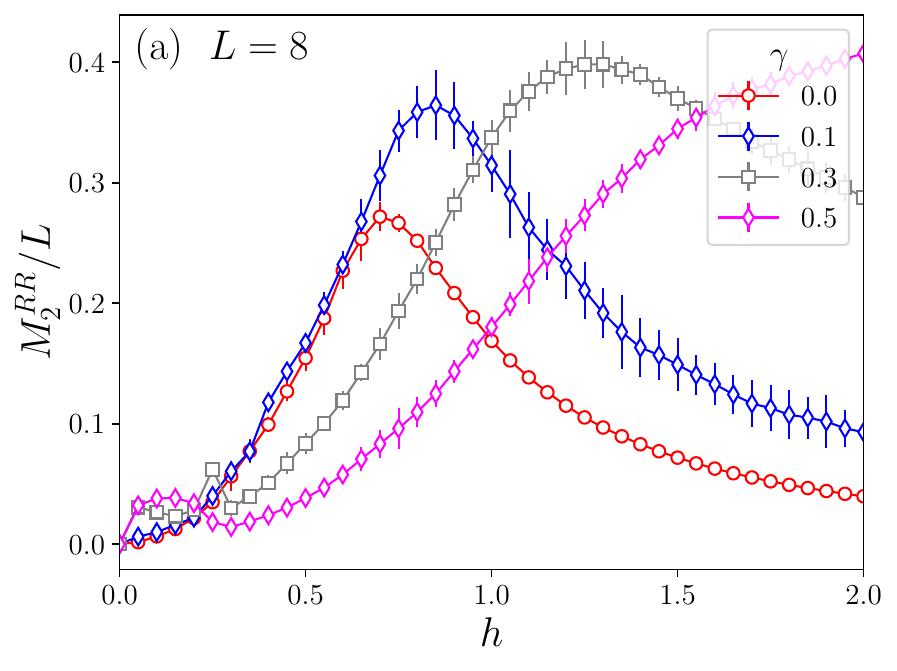}
		\includegraphics[width=0.95\columnwidth]{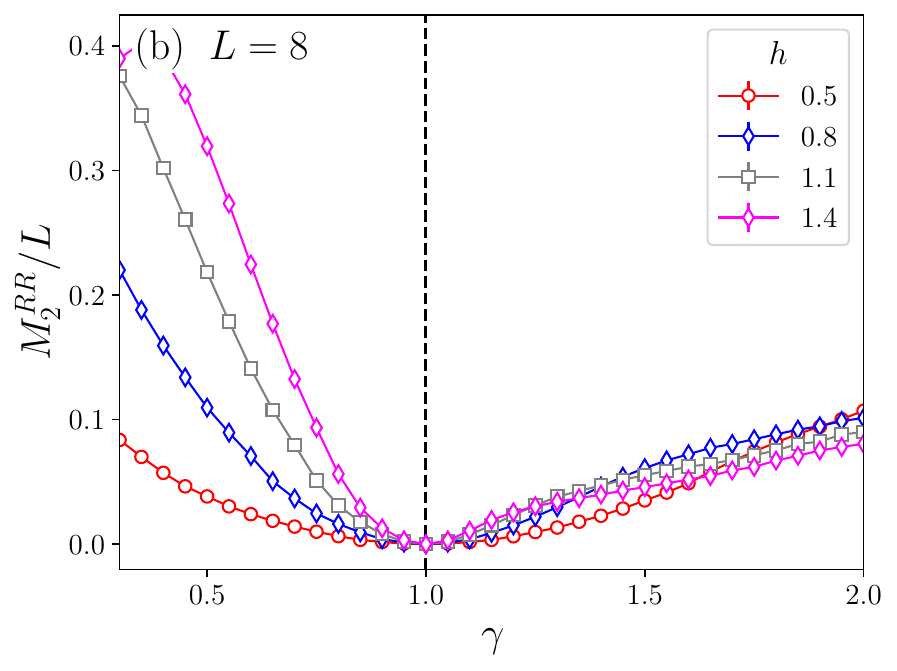}
		\caption{(a) Horizontal cuts through the phase diagram at several values of $\gamma$ in the $\mathcal{PT}$-symmetric regime.
		(b) Vertical cuts along the non-Hermitian parameter $\gamma$ for different external magnetic fields, illustrating that magic
		vanishes at the exceptional point $\gamma=1$.}
		\label{fig:Magic_h}
	\end{center}
\end{figure}

The stabilizer $2$-Rényi magic $M_2^{RR}$ in the ground state exhibits distinct features that reflect the
underlying phase structure of the model.
As shown in Fig.~\ref{fig:phase_diagram}, $M_2^{RR}$ peaks along the Ising transition line, indicating enhanced nonstabilizerness at the quantum critical point,
and vanishes at the exceptional points where the system undergoes a $\mathcal{PT}$-symmetry-breaking transition.
The position of the maximum in $M_2^{RR}$ shows a slight deviation from the exact critical line $\gamma_{c1}$.

In contrast, at the $\mathcal{PT}$ transition, $M_2^{RR}$ vanishes exactly, independent of the system size.
This can be understood from the similarity transformation~\cite{lu2024many},
which consists of a rotation around $\sigma^x$ and maps the non-Hermitian Hamiltonian onto a Hermitian one.
The rotation results in a Hermitian
transverse field Ising chain with an effective transverse field as $h\sqrt{1-\gamma^2}$. 
At the exceptional point with $\gamma=1$, the transverse field vanishes in the effective Hermitian model and it reduces to a simple Ising chain.
Its ground state is a fully polarized ferromagnet with $\langle \sigma_j^x\rangle=\pm 1$. The ground state wavefunction of the non-Hermitian model is
obtained by an inverse similarity transformation on the wavefunction, which means an inverse rotation around $\sigma^x$. However, since the Hermitian ground state is fully polarized in the $x$
direction in spin space, it remains intact by the inverse similarity transformation and the ground state of the non-Hermitian Ising model along the exceptional line is the fully ferromagnetic
state.
This is a trivial stabilizer product state and its magic is zero.

By analyzing horizontal and vertical cuts through the phase diagram (Fig.~\ref{fig:Magic_h}), one clearly observes that magic reaches its maximum at the critical line $\gamma_{c1}$ separating the ferromagnetic and paramagnetic phases, while it drops to zero at the full $\mathcal{PT}$-transition line
$\gamma_{c2} = 1$.
\begin{figure}[th!]
	\begin{center}
		\includegraphics[width=0.95\columnwidth]{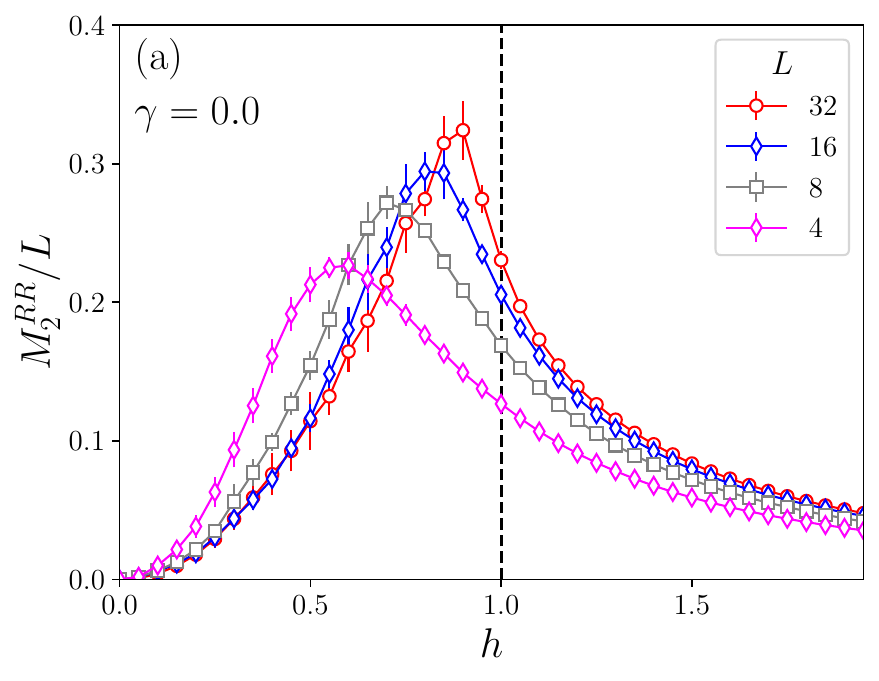}
		\includegraphics[width=0.95\columnwidth]{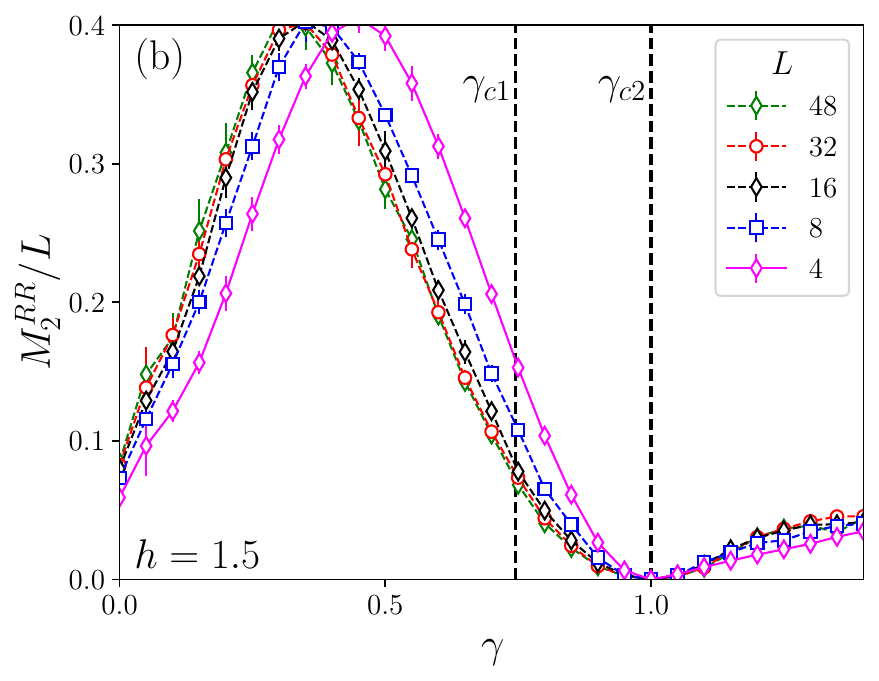}
		\caption{Scaling of $M_2^{RR}$ with system size in the non-Hermitian Ising model at (a) fixed $\gamma=0$ and (b) fixed $h=1.5$. 
		(a) As the system size increases, the location at the transition at $\gamma_{c1}$ becomes sharper and better resolved, while in (b), at fixed $h=1.5$, a shift of the maximum of $M_2^{RR}$ from $\gamma_{c1}$ is observed. 
		The $\mathcal{PT}$-symmetry transition at $\gamma_{c2}=1$ remains sharp and size independent.
		}
		\label{fig:Magic_scaling}
	\end{center}
\end{figure}

The behavior of $M_2^{RR}$ as a function of system size provides additional insight. 
As shown in Fig.~\ref{fig:Magic_scaling}(a) at $\gamma=0$, increasing the chain length sharpens the features of $M_2^{RR}$ near the critical lines, making the peaks at the Ising transition more pronounced. 
This scaling indicates that the maximum of $M_2^{RR}$ converges toward the true critical line $\gamma_{c1}$ in the thermodynamic limit.
Nevertheless, as shown in Fig.~\ref{fig:Magic_scaling}(b) at fixed $h=1.5$, there is a systematic slight shift of the maximum of $M_2^{RR}$ from the exact critical line $\gamma_{c1}$. In contrast, the drop to zero at $\gamma_{c2} = 1$ remains sharp and size independent, reflecting the exact vanishing of magic at the exceptional point.

This behavior demonstrates that $M_2^{RR}$ provides a direct and sensitive probe of non-Hermitian exceptional points, and approximates the position of the conventional Ising quantum phase transition
allowing one to partially reconstruct the ground-state phase diagram of the system from the pattern of magic alone.

\section{Non-Hermitian XX model}

\begin{figure}[t]
	\begin{center}
		\includegraphics[width=0.85\columnwidth]{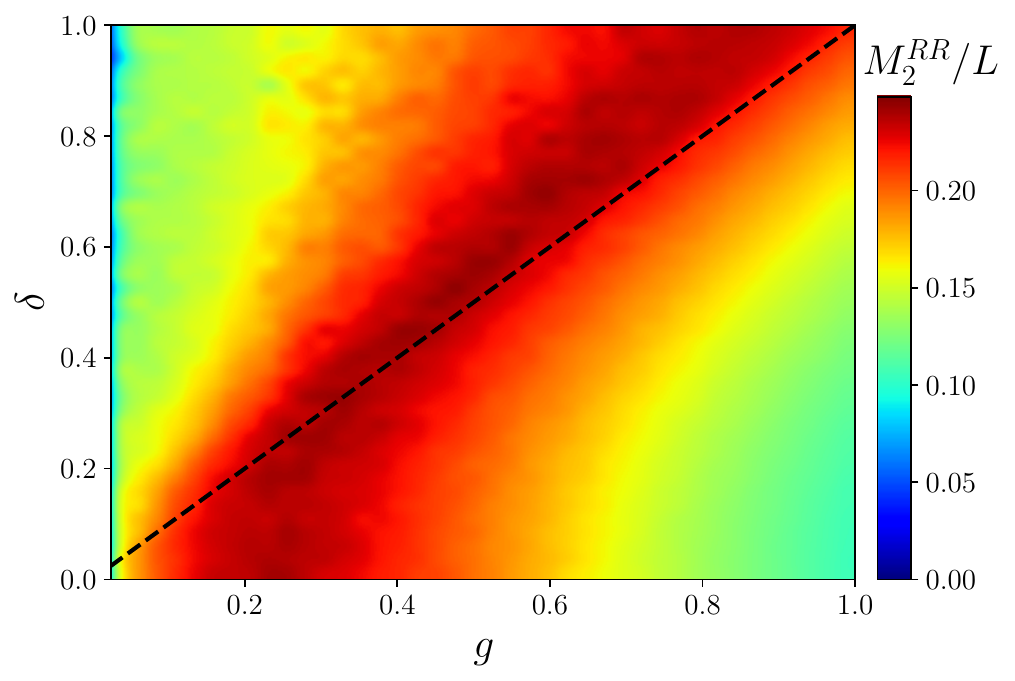}
		\caption{Ground-state density plot of $M_2^{RR}$ in the non-Hermitian XX model.
		The dashed black lines indicate the theoretical exceptional points line.
		Magic peaks along the $\mathcal{PT}$ exceptional line.
		The system size is $L=16$.}
		\label{fig:XX_phase_diagram}
	\end{center}
\end{figure}

The non-Hermitian XX model provides a complementary setting to investigate the interplay between $\mathcal{PT}$ symmetry and nonstabilizerness.
We contrast here the magic computed in real space lattices with OBC, and momentum-space magic~\cite{Dora2024}.
In both cases, nonstabilizerness takes extremal values at the exceptional points in the phase diagram.

\subsection{Real-space nonstabilizerness}

The starting point is a non-Hermitian tight-binding model for spinless fermions~\cite{Ashida2018,dora2022correlations}, endowed with $\mathcal{PT}$ symmetry, and described by the Hamiltonian
\begin{equation}
	H = \sum_{j=1}^{L-1} [J + i (-1)^j \delta]( c_j^\dagger c_{j+1} + \text{H.c.}) + 2g\sum_{j=1}^L (-1)^j c_j^\dagger c_j,
	\label{eq:H_fermions}
\end{equation}
with $J$, the uniform exchange coupling, $\delta$, a staggered non-Hermitian component, and $g$, a staggered longitudinal field.
Diagonalization yields a two-band spectrum,
\begin{equation}
	E_\pm(k) = \pm 2\sqrt{(J^2+\delta^2)\cos^2 k + g^2 - \delta^2},
\end{equation}
with momenta $k\in[0,\pi)$.
For $g > \delta$, the spectrum is real, and the system lies in the $\mathcal{PT}$-symmetric phase.
At the critical line $g=\delta$, the bands touch at $k=\pi/2$, signaling a
$\mathcal{PT}$-symmetric exceptional point with linear dispersion.
At half filling, the exceptional points lies at the Fermi level.
For $g < \delta$, the spectrum becomes complex near $k=\pi/2$, indicating entry into the $\mathcal{PT}$-broken regime where eigenvalues occur in
conjugate pairs.

%This Hamiltonian is obtained from a staggered tight-binding model of spinless fermions~\cite{Ashida2018,dora2022correlations} via Jordan-Wigner transformation, and preserves $\mathcal{PT}$ symmetry despite the complex exchange amplitude.

In order to study nonstabilizerness, we map the fermionic Hamiltonian to a spin (qubit) model.
Since the mapping represents a basis change and nonstabilizerness is basis dependent, the resulting magic should not be interpreted as a property of the original fermionic model. 
Instead, the ensuing spin models should be understood as physically motivated encodings, where magic is studied in a fixed computational basis.
Under a Jordan-Wigner transformation, Eq.~\eqref{eq:H_fermions} maps to a non-Hermitian XX chain with staggered complex exchange and staggered transverse field~\cite{Ashida2017},
% \begin{align}
% H_{\text{XX}} &= \sum_{j=1}^{L-1} \frac{J + i (-1)^j \delta}{2}\left( \sigma_j^x \sigma_{j+1}^x + \sigma_j^y \sigma_{j+1}^y \right)\nonumber \\
% &\quad + g\sum_{j=1}^L (-1)^j \sigma_j^z.
% \label{eq:HXX}
% \end{align}
\begin{align}
	H_{\text{XX}} & = \sum_{j=1}^{L-1} [J + i (-1)^j \delta]( \sigma_j^+ \sigma_{j+1}^- + \text{H.c.})\notag \\
	              & \quad + g\sum_{j=1}^L (-1)^j \sigma_j^z,
	\label{eq:HXX}
\end{align}
with $\sigma^\pm=\frac12(\sigma^x\pm i\sigma^y)$.
Its phases remain the same: a gapped $\mathcal{PT}$-symmetric phase, a critical line of exceptional points at $g=\delta$, and a $\mathcal{PT}$-broken phase with complex spectrum.
This structure parallels that of the non-Hermitian Ising chain, and provides an alternative platform where stabilizer R\'enyi entropies can be employed to probe both conventional criticality and non-Hermitian exceptional points.

\begin{figure}[th!]
	\begin{center}
		\includegraphics[width=0.95\columnwidth]{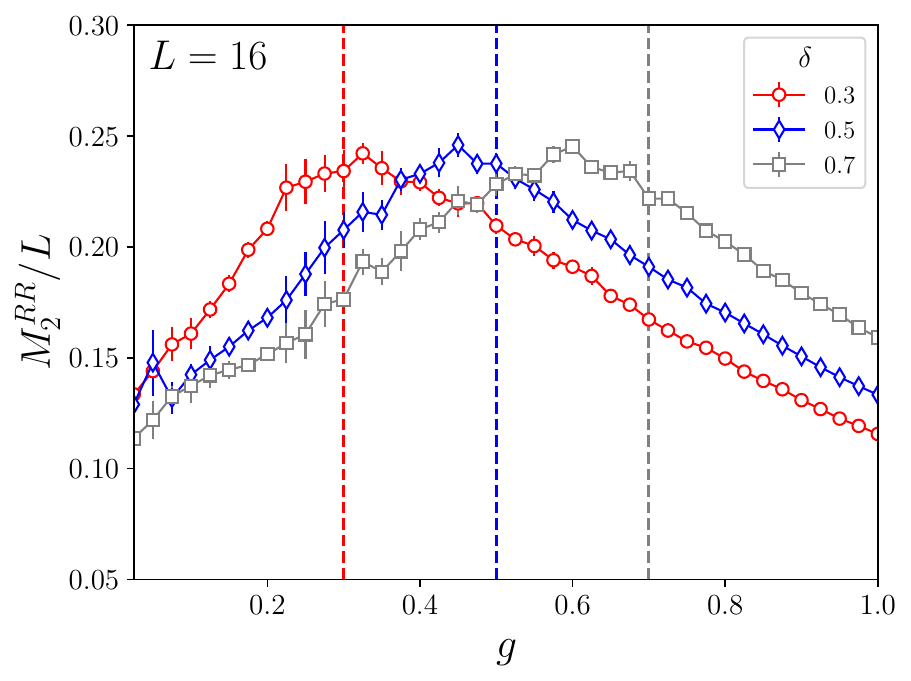}
		\caption{Horizontal cuts through the phase diagram of the non-Hermitian XX model at several values of $\delta$ showing that $M_2^{RR}$ peaks at the exceptional point $g=\delta$, marked by dashed lines of the corresponding color. }
		\label{fig:XX_Magic_h}
	\end{center}
\end{figure}

The stabilizer $2$-R\'enyi magic $M_2^{RR}$ in the ground state of the XX model displays distinctive signatures that mirror this phase structure.
As shown in Fig.~\ref{fig:XX_phase_diagram}, $M_2^{RR}$ develops pronounced ridges along the exceptional line $g=\delta$, reaching its maximum value close to the $\mathcal{PT}$-symmetric exceptional points.
In contrast to the transverse-field Ising case, where magic vanishes at the exceptional transition, here the nonstabilizerness is maximized at the onset of $\mathcal{PT}$-symmetry breaking.
This highlights the model dependence of magic as a diagnostic: While in the Ising chain it signals conventional quantum criticality, in the XX model it is  sensitive to exceptional points.
Horizontal cuts through the phase diagram (Fig.~\ref{fig:XX_Magic_h}) confirm this behavior, showing peaks of $M_2^{RR}$
pinned to the exceptional line.

\subsection{Momentum space nonstabilizerness}
A complementary perspective on the non-Hermitian model~\eqref{eq:H_fermions} consists in determining its nonstabilizerness in momentum space, $\mathcal M^{RR}_2$~\cite{Dora2024}. This originates from the proposal in Ref. ~\cite{verstraete2009}, which investigated the simulation of strongly correlated quantum many-body systems on finite-depth quantum circuits that diagonalize their dynamics, which can involve
working in momentum space. This was carried out on a quantum devices~\cite{Cervera} by focusing on the momentum space simulation of the transverse field quantum Ising chain. These ideas are now applied to the XX chain. 
The XX model becomes analytically solvable in momentum space, where it is diagonalized in separate $k$ sectors, and the total magic becomes a trivial sum of few-qubit results from each sector.
To determine a momentum-space spin Hamiltonian, we start with the fermionic Hamiltonian~\eqref{eq:H_fermions} with periodic boundary conditions, in momentum space,
% \begin{equation}\label{eq:H_k}
% H = 2\sum_{k\in[0,\pi)}C^\dag_k
% \begin{pmatrix}
% J\cos k & g+\delta \sin k \\
% g-\delta \sin k & -J\cos k
% \end{pmatrix}
% C_k,
% \end{equation}
\begin{equation}\label{eq:H_k}
	H = \sum_{k=0}^\pi C^\dag_k
	H_k
	C_k,\; \frac{H_k}{2}=\begin{pmatrix}
		J\cos k         & g+\delta \sin k \\
		g-\delta \sin k & -J\cos k
	\end{pmatrix},
\end{equation}
with $C^\dag_k=(c^\dag_k,c^\dag_{k-\pi})$.
Then, we perform a Jordan-Wigner transformation, mapping fermion to spin operators. For example, the annihilation operators are represented as
\begin{equation}
	c_k=\exp\left[i\frac{\pi}{2}\sum_{q\in\mathcal{D}_k}(\sigma^z_q+\sigma^0_q)\right]\sigma^-_k,
\end{equation}
with momenta $q$ in the set $\mathcal{D}_k=[-\pi,k-\pi)\cup[0,k)$.
Thus, the number operator is $c_k^\dag c_k^{}=\sigma^+_k\sigma^-_k\equiv\frac{1}{2}(\sigma^z_k+\sigma^0_k)$, with $\sigma^0_k$ the $2\times 2$ identity matrix.
Inside the Jordan-Wigner string, the order of momenta is in pairs $(q-\pi,q)$, such that the Jordan-Wigner transformations reads explicitly
\begin{equation}
	c_k = \left(\prod_{q\in[0,k)}\sigma^z_{q-\pi}\sigma^z_q\right)\sigma^-_k,
\end{equation}
with momenta $q$ increasing from $0$ to $k$.
The transformation yields the momentum space Hamiltonian
\begin{widetext}
	\begin{equation}
		H = \sum_{k\in[0,\pi)}J\cos k(\sigma^z_k-\sigma^z_{k-\pi})+2(g+\delta\sin k)\sigma^+_k\sigma_{k-\pi}^-
		+2(g-\delta\sin k)\sigma^-_k\sigma_{k-\pi}^+.
		\label{eq:H_k_spin}
	\end{equation}
\end{widetext}
Since only $k$ and $k-\pi$ momenta remain coupled in this Hamiltonian, it is meaningful to represent it in the two-qubit basis
$\{|0\rangle_{k-\pi}\otimes|0\rangle_k,
	|1\rangle_{k-\pi}\otimes|1\rangle_k,
	|0\rangle_{k-\pi}\otimes|1\rangle_k,
	|1\rangle_{k-\pi}\otimes|0\rangle_k\}$.
To simplify the notation, we denote $|0\rangle_{k-\pi}\otimes|0\rangle_k$ as $|00\rangle_k$, etc.
The qubit states are defined such that $\sigma^z_k|0\rangle_k=-|0\rangle_k$ and $\sigma^z_k|1\rangle_k=|1\rangle_k$.
Since the Hamiltonian conserves the number of spin excitations, it assumes for each momentum $k$ a $2\times 2$ matrix form in the reduced basis $\{|01\rangle_k,|10\rangle_k\}$, where it is represented by $H_k$ from Eq.~\eqref{eq:H_k}.

\begin{figure}[t]
	\centering
	\includegraphics[width=0.8\columnwidth]{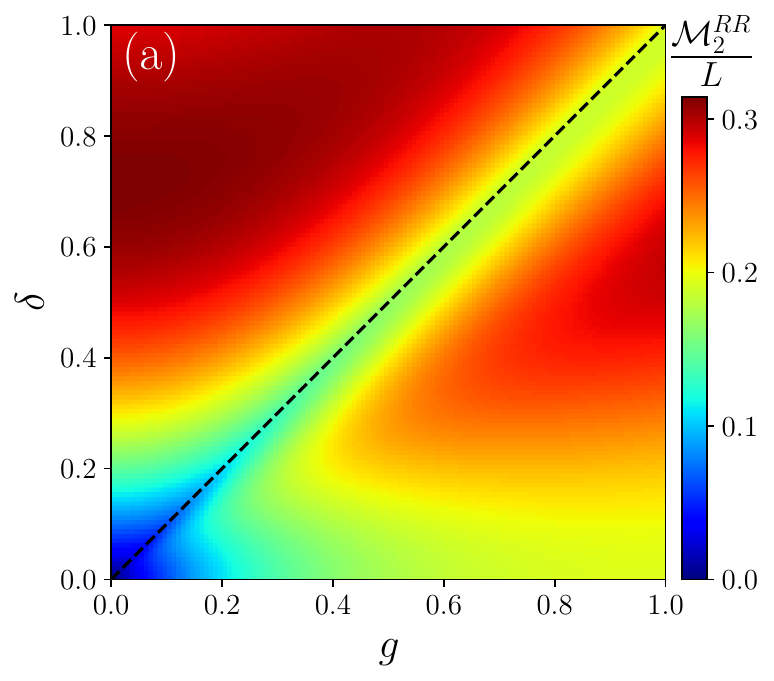}
	\includegraphics[width=0.8\columnwidth]{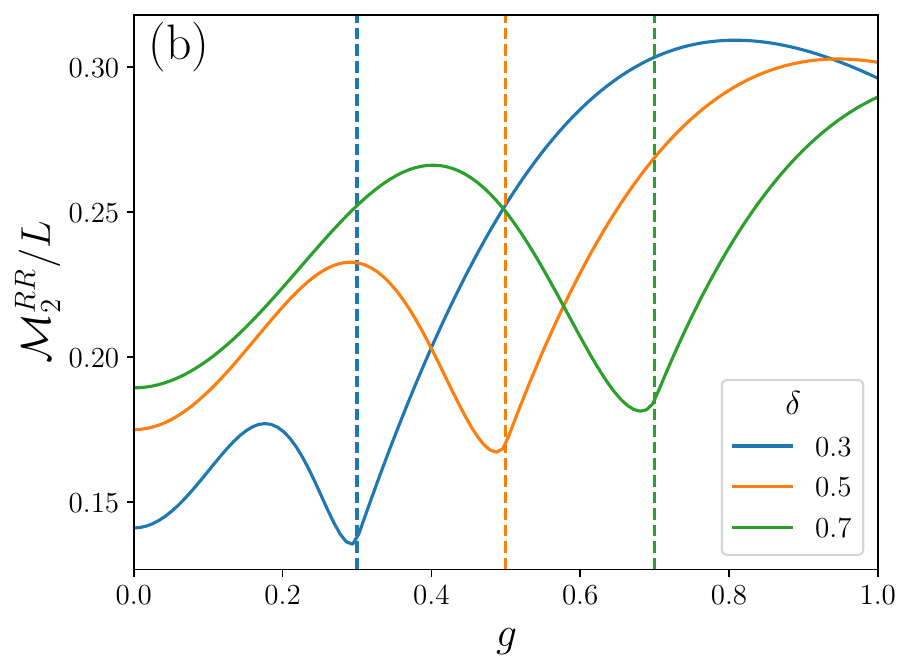}
	\caption{(a) Ground-state magic density $\mathcal{M}^{RR}_2/L$ in momentum space for the XX model, using a mesh of $L=800$ momenta in $[0,\pi)$.
	The dashed line indicates the exceptional line in parameter space.
	(b) Horizontal cuts through the phase diagram at several values of $\delta$ showing that $\mathcal{M}_2^{RR}/L$ has a local minimum at the exceptional points marked by dashed lines of the corresponding color.}
	\label{fig:XX_Magic_k}
\end{figure}

The ground state of the system is a product state in momentum space
$|\Phi\rangle = \bigotimes_{k\in[0,\pi)}|\phi_k\rangle$, with
\begin{equation}
	|\phi_k\rangle= \alpha_k |0 1\rangle_k
	+ \beta_k |10\rangle_k,
\end{equation}
and is the state with the lowest real energy of $H_k$ for each momentum $k$.
In this sense, the ground state is a classical state with no entanglement between different momentum sectors.
% The ground state is defined as the state with the lowest real energy of $H_k$ for each momentum $k$.
Then, its nonstabilizerness depends entirely on the coefficients $\alpha_k$ and $\beta_k$, and the total magic is obtained by summing the contributions from each momentum sector.
The resulting diagram for the magic density $\mathcal{M}_2^{RR}/L$ in $g$--$\delta$ plane are shown in Fig.~\ref{fig:XX_Magic_k}(a), where $L$ is the number of momenta in the Brillouin zone $[0,\pi)$.
Additionally, a few cuts through the phase diagram at fixed $\delta$ are shown in Fig.~\ref{fig:XX_Magic_k}(b) illustrating that at exceptional points, nonstabilizerness reaches a local minimum.
As expected, near the $g=\delta=0$ point, the magic density vanishes, since the ground state is a stabilizer state, a product of eigenstates of $\sigma^z$, as evident from Eq.~\eqref{eq:H_k_spin}.
Moving along the line of exceptional points $g=\delta$, the magic density increases until it saturates at some finite value.
Solving Eq.~\eqref{eq:H_k} for the ground state $\psi_R$ on the exceptional line, we find
\begin{equation}
\psi_R=\frac{1}{N}
\begin{pmatrix}-g(1+\sin k)\\ J\cos k+\sqrt{J^2+g^2}|\cos k|\end{pmatrix},
\end{equation}
where $N$ normalizes the state.
Thus, on the exceptional line, the qubit in each momentum sector rotates in the $x$-$z$ plane, simplifying the analysis.
The saturation value on the exceptional line is found analytically by taking the limit $g=\delta\to\infty$ for $\psi_R$ and computing the magic using Eq.~\eqref{eq:M2RR},
\begin{align}\label{eq:magic_limit}
	\lim_{g\to\infty}\frac{\mathcal{M}_2^{RR}}{L} & =-\frac{1}{\pi}\int_0^\pi dk \log_2\frac{1+\sin^4k+\cos^4k}{2}\notag \\
	& =\log_2(112-64\sqrt3)\approx 0.2.
\end{align}

\begin{figure}[t]
	\includegraphics[width=0.8\columnwidth]{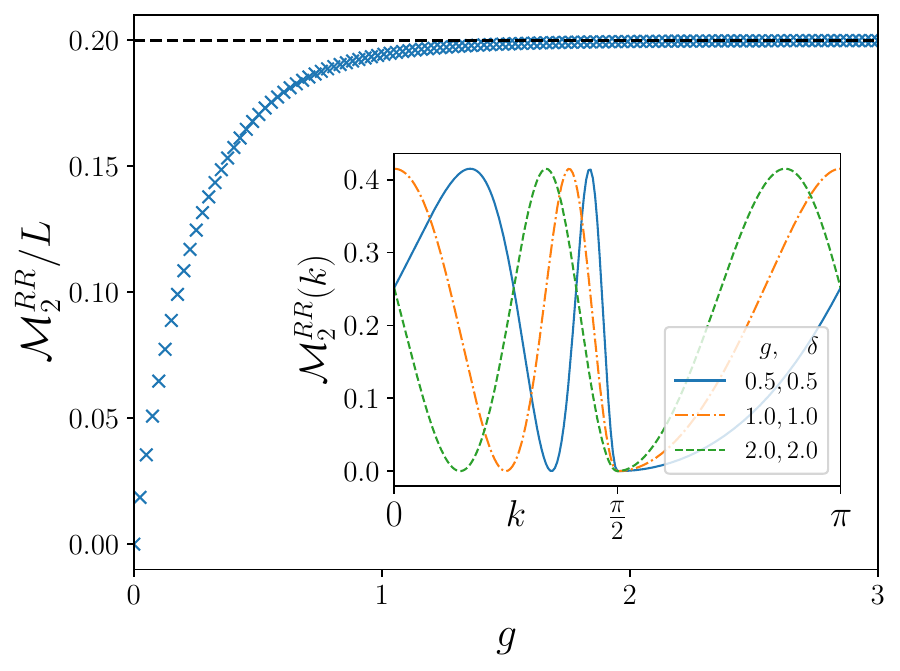}
	\caption{Magic density along the exceptional line $g=\delta$ rapidly saturates to the (dashed line) analytical limit~\eqref{eq:magic_limit}.
		The inset shows the momentum-resolved magic for several values on the exceptional line. The magic is zero at the exceptional point $k=\pi/2$, while an additional magic minimum occurs at $k<\pi/2$ when the qubit orients itself along the $x$ axis.
	}
	\label{fig:magic_saturation}
\end{figure}

In Fig.~\ref{fig:magic_saturation}, we show the rapid convergence of magic density to the analytical limit when increasing $g=\delta$.
Moreover, in the inset of Fig.~\ref{fig:magic_saturation}, we show the momentum-resolved magic for several values on the exceptional line, demonstrating that exactly at the exceptional point $k=\pi/2$, the magic vanishes, as the qubit aligns along the $z$ axis.
Interestingly, as the qubit state rotates in the $x$-$z$ plane, it passes through a second magic minimum, where it becomes a stabilizer state, being completely aligned along the $x$ axis.

Importantly, the magic density attains a minimum along the exceptional line in the $g$-$\delta$ phase diagram.
This behavior contrasts sharply with that of the real-space magic, which instead peaks at the exceptional line.
Such a discrepancy arises from the basis dependence of magic.
Nonetheless, in both representations, the extreme values of magic reliably signal the onset of $\mathcal{PT}$-symmetry breaking at the exceptional points.

\section{Conclusions}

In this work, we have explored nonstabilizerness, quantified by stabilizer R\'enyi entropies, as a diagnostic tool for non-Hermitian quantum many-body systems.
Using the non-Hermitian transverse-field Ising chain and the
non-Hermitian XX chain as models, we analyzed ground-state properties across their respective phase diagrams.

Our results show that magic provides a clear signature of  the quantum phase transitions in non-Hermitian systems.
In the Ising chain, $M_2^{RR}$ peaks near the critical line, signaling the presence of non-classical resources at the transition, and vanishes exactly across exceptional points, reflecting the coalescence of eigenstates.

By contrast, in the real-space XX chain $M_2^{RR}$ is maximized along the exceptional line.
We have also analyzed magic in momentum space for the XX model and found that it reaches a minimum at exceptional points, in contrast to its real space counterpart. This clearly highlights the essential basis dependence of the magic.

Finite-size scaling further confirms that these features sharpen with increasing system size, establishing $M_2^{RR}$ as a reliable probe of non-Hermitian criticality.

Taken together, our findings demonstrate that nonstabilizerness captures essential equilibrium features of non-Hermitian quantum matter in ways that
complement traditional measures such as entanglement entropy.
Magic thus emerges as a versatile addition to the toolbox for characterizing resource properties in non-Hermitian systems.

Looking ahead, it will be interesting to extend these ideas to disordered or non-equilibrium systems and experimental platforms where non-Hermitian physics and quantum information concepts meet~\cite{Kattel2023,Turkeshi2025,Tirrito2025a,sticlet2025non}.

\section{Acknowledgments}
This work received financial support from CNCS/CCCDI-UEFISCDI, 
under Projects No.~PN-IV-P1-PCE-2023-0159, and No.~PN-IV-P1-PCE-2023-0987, and by the ``Nucleu'' Program within the PNCDI 2022-2027, Romania, carried out with the support of MEC, Project No.~27N/03.01.2023, component project code PN 23 24 01 04.
This research was also supported by the National Research, Development and Innovation Office - NKFIH within the Quantum Technology National Excellence
Program (Project No. 2017-1.2.1-NKP-2017-00001), K134437, K142179 by the BME-Nanotechnology
FIKP grant (BME FIKP-NAT).
We acknowledge the Digital Government Development and Project Management Ltd.~for awarding us access to the Komondor HPC facility based in Hungary.

\bibliography{references}

%apsrev4-2.bst 2019-01-14 (MD) hand-edited version of apsrev4-1.bst
%Control: key (0)
%Control: author (8) initials jnrlst
%Control: editor formatted (1) identically to author
%Control: production of article title (0) allowed
%Control: page (0) single
%Control: year (1) truncated
%Control: production of eprint (0) enabled
\begin{thebibliography}{72}%
\makeatletter
\providecommand \@ifxundefined [1]{%
 \@ifx{#1\undefined}
}%
\providecommand \@ifnum [1]{%
 \ifnum #1\expandafter \@firstoftwo
 \else \expandafter \@secondoftwo
 \fi
}%
\providecommand \@ifx [1]{%
 \ifx #1\expandafter \@firstoftwo
 \else \expandafter \@secondoftwo
 \fi
}%
\providecommand \natexlab [1]{#1}%
\providecommand \enquote  [1]{``#1''}%
\providecommand \bibnamefont  [1]{#1}%
\providecommand \bibfnamefont [1]{#1}%
\providecommand \citenamefont [1]{#1}%
\providecommand \href@noop [0]{\@secondoftwo}%
\providecommand \href [0]{\begingroup \@sanitize@url \@href}%
\providecommand \@href[1]{\@@startlink{#1}\@@href}%
\providecommand \@@href[1]{\endgroup#1\@@endlink}%
\providecommand \@sanitize@url [0]{\catcode `\\12\catcode `\$12\catcode
  `\&12\catcode `\#12\catcode `\^12\catcode `\_12\catcode `\%12\relax}%
\providecommand \@@startlink[1]{}%
\providecommand \@@endlink[0]{}%
\providecommand \url  [0]{\begingroup\@sanitize@url \@url }%
\providecommand \@url [1]{\endgroup\@href {#1}{\urlprefix }}%
\providecommand \urlprefix  [0]{URL }%
\providecommand \Eprint [0]{\href }%
\providecommand \doibase [0]{https://doi.org/}%
\providecommand \selectlanguage [0]{\@gobble}%
\providecommand \bibinfo  [0]{\@secondoftwo}%
\providecommand \bibfield  [0]{\@secondoftwo}%
\providecommand \translation [1]{[#1]}%
\providecommand \BibitemOpen [0]{}%
\providecommand \bibitemStop [0]{}%
\providecommand \bibitemNoStop [0]{.\EOS\space}%
\providecommand \EOS [0]{\spacefactor3000\relax}%
\providecommand \BibitemShut  [1]{\csname bibitem#1\endcsname}%
\let\auto@bib@innerbib\@empty
%</preamble>
\bibitem [{\citenamefont {Moiseyev}(2011)}]{moiseyev2011non}%
  \BibitemOpen
  \bibfield  {author} {\bibinfo {author} {\bibfnamefont {N.}~\bibnamefont
  {Moiseyev}},\ }\href {https://doi.org/10.1017/CBO9780511976186} {\emph
  {\bibinfo {title} {Non-{Hermitian} Quantum Mechanics}}}\ (\bibinfo
  {publisher} {Cambridge University Press},\ \bibinfo {address} {Cambridge},\
  \bibinfo {year} {2011})\BibitemShut {NoStop}%
\bibitem [{\citenamefont {Ashida}\ \emph {et~al.}(2020)\citenamefont {Ashida},
  \citenamefont {Gong},\ and\ \citenamefont {Ueda}}]{ashida2020non}%
  \BibitemOpen
  \bibfield  {author} {\bibinfo {author} {\bibfnamefont {Y.}~\bibnamefont
  {Ashida}}, \bibinfo {author} {\bibfnamefont {Z.}~\bibnamefont {Gong}},\ and\
  \bibinfo {author} {\bibfnamefont {M.}~\bibnamefont {Ueda}},\ }\bibfield
  {title} {\bibinfo {title} {Non-{Hermitian} physics},\ }\href
  {https://doi.org/10.1080/00018732.2021.1876991} {\bibfield  {journal}
  {\bibinfo  {journal} {Adv. Phys.}\ }\textbf {\bibinfo {volume} {69}},\
  \bibinfo {pages} {249–435} (\bibinfo {year} {2020})}\BibitemShut {NoStop}%
\bibitem [{\citenamefont {Wang}\ \emph {et~al.}(2023)\citenamefont {Wang},
  \citenamefont {Fu}, \citenamefont {Mao}, \citenamefont {Qie}, \citenamefont
  {Stone},\ and\ \citenamefont {Yang}}]{wang2023non}%
  \BibitemOpen
  \bibfield  {author} {\bibinfo {author} {\bibfnamefont {C.}~\bibnamefont
  {Wang}}, \bibinfo {author} {\bibfnamefont {Z.}~\bibnamefont {Fu}}, \bibinfo
  {author} {\bibfnamefont {W.}~\bibnamefont {Mao}}, \bibinfo {author}
  {\bibfnamefont {J.}~\bibnamefont {Qie}}, \bibinfo {author} {\bibfnamefont
  {A.~D.}\ \bibnamefont {Stone}},\ and\ \bibinfo {author} {\bibfnamefont
  {L.}~\bibnamefont {Yang}},\ }\bibfield  {title} {\bibinfo {title}
  {Non-{Hermitian} optics and photonics: {From} classical to quantum},\ }\href
  {https://doi.org/10.1364/AOP.475477} {\bibfield  {journal} {\bibinfo
  {journal} {Adv. Opt. Photon.}\ }\textbf {\bibinfo {volume} {15}},\ \bibinfo
  {pages} {442} (\bibinfo {year} {2023})}\BibitemShut {NoStop}%
\bibitem [{\citenamefont {Rotter}(2009)}]{rotter2009non}%
  \BibitemOpen
  \bibfield  {author} {\bibinfo {author} {\bibfnamefont {I.}~\bibnamefont
  {Rotter}},\ }\bibfield  {title} {\bibinfo {title} {A non-{Hermitian}
  {Hamilton} operator and the physics of open quantum systems},\ }\href
  {https://doi.org/10.1088/1751-8113/42/15/153001} {\bibfield  {journal}
  {\bibinfo  {journal} {J. Phys. A: Math. Theor.}\ }\textbf {\bibinfo {volume}
  {42}},\ \bibinfo {pages} {153001} (\bibinfo {year} {2009})}\BibitemShut
  {NoStop}%
\bibitem [{\citenamefont {Roccati}\ \emph {et~al.}(2022)\citenamefont
  {Roccati}, \citenamefont {Palma}, \citenamefont {Ciccarello},\ and\
  \citenamefont {Bagarello}}]{roccati2022non}%
  \BibitemOpen
  \bibfield  {author} {\bibinfo {author} {\bibfnamefont {F.}~\bibnamefont
  {Roccati}}, \bibinfo {author} {\bibfnamefont {G.~M.}\ \bibnamefont {Palma}},
  \bibinfo {author} {\bibfnamefont {F.}~\bibnamefont {Ciccarello}},\ and\
  \bibinfo {author} {\bibfnamefont {F.}~\bibnamefont {Bagarello}},\ }\bibfield
  {title} {\bibinfo {title} {Non-{Hermitian} physics and master equations},\
  }\href {https://doi.org/10.1142/S1230161222500044} {\bibfield  {journal}
  {\bibinfo  {journal} {Open Syst. Inf. Dyn.}\ }\textbf {\bibinfo {volume}
  {29}},\ \bibinfo {pages} {2250004} (\bibinfo {year} {2022})}\BibitemShut
  {NoStop}%
\bibitem [{\citenamefont {G\'omez-Le\'on}\ \emph {et~al.}(2022)\citenamefont
  {G\'omez-Le\'on}, \citenamefont {Ramos}, \citenamefont {Gonz\'alez-Tudela},\
  and\ \citenamefont {Porras}}]{gomez2022bridging}%
  \BibitemOpen
  \bibfield  {author} {\bibinfo {author} {\bibfnamefont {A.}~\bibnamefont
  {G\'omez-Le\'on}}, \bibinfo {author} {\bibfnamefont {T.}~\bibnamefont
  {Ramos}}, \bibinfo {author} {\bibfnamefont {A.}~\bibnamefont
  {Gonz\'alez-Tudela}},\ and\ \bibinfo {author} {\bibfnamefont
  {D.}~\bibnamefont {Porras}},\ }\bibfield  {title} {\bibinfo {title} {Bridging
  the gap between topological non-{Hermitian} physics and open quantum
  systems},\ }\href {https://doi.org/10.1103/PhysRevA.106.L011501} {\bibfield
  {journal} {\bibinfo  {journal} {Phys. Rev. A}\ }\textbf {\bibinfo {volume}
  {106}},\ \bibinfo {pages} {L011501} (\bibinfo {year} {2022})}\BibitemShut
  {NoStop}%
\bibitem [{\citenamefont {Yan}\ \emph {et~al.}(2023)\citenamefont {Yan},
  \citenamefont {Zhao}, \citenamefont {Zhou}, \citenamefont {Ma}, \citenamefont
  {Lyu}, \citenamefont {Chu}, \citenamefont {Hu},\ and\ \citenamefont
  {Gong}}]{yan2023advances}%
  \BibitemOpen
  \bibfield  {author} {\bibinfo {author} {\bibfnamefont {Q.}~\bibnamefont
  {Yan}}, \bibinfo {author} {\bibfnamefont {B.}~\bibnamefont {Zhao}}, \bibinfo
  {author} {\bibfnamefont {R.}~\bibnamefont {Zhou}}, \bibinfo {author}
  {\bibfnamefont {R.}~\bibnamefont {Ma}}, \bibinfo {author} {\bibfnamefont
  {Q.}~\bibnamefont {Lyu}}, \bibinfo {author} {\bibfnamefont {S.}~\bibnamefont
  {Chu}}, \bibinfo {author} {\bibfnamefont {X.}~\bibnamefont {Hu}},\ and\
  \bibinfo {author} {\bibfnamefont {Q.}~\bibnamefont {Gong}},\ }\bibfield
  {title} {\bibinfo {title} {Advances and applications on non-{Hermitian}
  topological photonics},\ }\href {https://doi.org/10.1515/nanoph-2022-0775}
  {\bibfield  {journal} {\bibinfo  {journal} {Proc. Spie.}\ }\textbf {\bibinfo
  {volume} {12}},\ \bibinfo {pages} {2247} (\bibinfo {year}
  {2023})}\BibitemShut {NoStop}%
\bibitem [{\citenamefont {Li}\ \emph {et~al.}(2023)\citenamefont {Li},
  \citenamefont {Wei}, \citenamefont {Cotrufo}, \citenamefont {Chen},
  \citenamefont {Mann}, \citenamefont {Ni}, \citenamefont {Xu}, \citenamefont
  {Chen}, \citenamefont {Wang}, \citenamefont {Fan}, \citenamefont {Qiu},
  \citenamefont {Alù},\ and\ \citenamefont {Chen}}]{li2023exceptional}%
  \BibitemOpen
  \bibfield  {author} {\bibinfo {author} {\bibfnamefont {A.}~\bibnamefont
  {Li}}, \bibinfo {author} {\bibfnamefont {H.}~\bibnamefont {Wei}}, \bibinfo
  {author} {\bibfnamefont {M.}~\bibnamefont {Cotrufo}}, \bibinfo {author}
  {\bibfnamefont {W.}~\bibnamefont {Chen}}, \bibinfo {author} {\bibfnamefont
  {S.}~\bibnamefont {Mann}}, \bibinfo {author} {\bibfnamefont {X.}~\bibnamefont
  {Ni}}, \bibinfo {author} {\bibfnamefont {B.}~\bibnamefont {Xu}}, \bibinfo
  {author} {\bibfnamefont {J.}~\bibnamefont {Chen}}, \bibinfo {author}
  {\bibfnamefont {J.}~\bibnamefont {Wang}}, \bibinfo {author} {\bibfnamefont
  {S.}~\bibnamefont {Fan}}, \bibinfo {author} {\bibfnamefont {C.-W.}\
  \bibnamefont {Qiu}}, \bibinfo {author} {\bibfnamefont {A.}~\bibnamefont
  {Alù}},\ and\ \bibinfo {author} {\bibfnamefont {L.}~\bibnamefont {Chen}},\
  }\bibfield  {title} {\bibinfo {title} {Exceptional points and non-{Hermitian}
  photonics at the nanoscale},\ }\href
  {https://doi.org/10.1038/s41565-023-01408-0} {\bibfield  {journal} {\bibinfo
  {journal} {Nat. Nanotechnol.}\ }\textbf {\bibinfo {volume} {18}},\ \bibinfo
  {pages} {706} (\bibinfo {year} {2023})}\BibitemShut {NoStop}%
\bibitem [{\citenamefont {Li}\ \emph {et~al.}(2020)\citenamefont {Li},
  \citenamefont {Lee},\ and\ \citenamefont {Gong}}]{li2020topological}%
  \BibitemOpen
  \bibfield  {author} {\bibinfo {author} {\bibfnamefont {L.}~\bibnamefont
  {Li}}, \bibinfo {author} {\bibfnamefont {C.~H.}\ \bibnamefont {Lee}},\ and\
  \bibinfo {author} {\bibfnamefont {J.}~\bibnamefont {Gong}},\ }\bibfield
  {title} {\bibinfo {title} {Topological switch for non-{Hermitian} skin effect
  in cold-atom systems with loss},\ }\href
  {https://doi.org/10.1103/PhysRevLett.124.250402} {\bibfield  {journal}
  {\bibinfo  {journal} {Phys. Rev. Lett.}\ }\textbf {\bibinfo {volume} {124}},\
  \bibinfo {pages} {250402} (\bibinfo {year} {2020})}\BibitemShut {NoStop}%
\bibitem [{\citenamefont {Luo}\ \emph {et~al.}(2024)\citenamefont {Luo},
  \citenamefont {Liu},\ and\ \citenamefont {Liang}}]{luo2024quantum}%
  \BibitemOpen
  \bibfield  {author} {\bibinfo {author} {\bibfnamefont {Y.}~\bibnamefont
  {Luo}}, \bibinfo {author} {\bibfnamefont {N.}~\bibnamefont {Liu}},\ and\
  \bibinfo {author} {\bibfnamefont {J.-Q.}\ \bibnamefont {Liang}},\ }\bibfield
  {title} {\bibinfo {title} {Quantum phase transition and exceptional points of
  a non-{Hermitian} {Hamiltonian} for cold atoms in a dissipative optical
  cavity with nonlinear atom-photon interactions},\ }\href
  {https://doi.org/10.1103/PhysRevA.110.063320} {\bibfield  {journal} {\bibinfo
   {journal} {Phys. Rev. A}\ }\textbf {\bibinfo {volume} {110}},\ \bibinfo
  {pages} {063320} (\bibinfo {year} {2024})}\BibitemShut {NoStop}%
\bibitem [{\citenamefont {Mostafazadeh}(2002)}]{mostafazadeh2002pseudo}%
  \BibitemOpen
  \bibfield  {author} {\bibinfo {author} {\bibfnamefont {A.}~\bibnamefont
  {Mostafazadeh}},\ }\bibfield  {title} {\bibinfo {title} {Pseudo-{Hermiticity}
  versus {PT} symmetry: {The} necessary condition for the reality of the
  spectrum of a non-{Hermitian} {Hamiltonian}},\ }\href
  {https://doi.org/10.1063/1.1418246} {\bibfield  {journal} {\bibinfo
  {journal} {J. Math. Phys.}\ }\textbf {\bibinfo {volume} {43}},\ \bibinfo
  {pages} {205–214} (\bibinfo {year} {2002})}\BibitemShut {NoStop}%
\bibitem [{\citenamefont {Mostafazadeh}(2003)}]{mostafazadeh2003exact}%
  \BibitemOpen
  \bibfield  {author} {\bibinfo {author} {\bibfnamefont {A.}~\bibnamefont
  {Mostafazadeh}},\ }\bibfield  {title} {\bibinfo {title} {Exact {PT}-symmetry
  is equivalent to {Hermiticity}},\ }\href
  {https://doi.org/10.1088/0305-4470/36/25/312} {\bibfield  {journal} {\bibinfo
   {journal} {J. Phys. A: Math. Gen.}\ }\textbf {\bibinfo {volume} {36}},\
  \bibinfo {pages} {7081} (\bibinfo {year} {2003})}\BibitemShut {NoStop}%
\bibitem [{\citenamefont {Feng}\ \emph {et~al.}(2017)\citenamefont {Feng},
  \citenamefont {El-Ganainy},\ and\ \citenamefont {Ge}}]{feng2017non}%
  \BibitemOpen
  \bibfield  {author} {\bibinfo {author} {\bibfnamefont {L.}~\bibnamefont
  {Feng}}, \bibinfo {author} {\bibfnamefont {R.}~\bibnamefont {El-Ganainy}},\
  and\ \bibinfo {author} {\bibfnamefont {L.}~\bibnamefont {Ge}},\ }\bibfield
  {title} {\bibinfo {title} {Non-{Hermitian} photonics based on parity--time
  symmetry},\ }\href {https://doi.org/10.1038/s41566-017-0031-1} {\bibfield
  {journal} {\bibinfo  {journal} {Nat. Photonics}\ }\textbf {\bibinfo {volume}
  {11}},\ \bibinfo {pages} {752} (\bibinfo {year} {2017})}\BibitemShut
  {NoStop}%
\bibitem [{\citenamefont {Longhi}(2018)}]{longhi2018parity}%
  \BibitemOpen
  \bibfield  {author} {\bibinfo {author} {\bibfnamefont {S.}~\bibnamefont
  {Longhi}},\ }\bibfield  {title} {\bibinfo {title} {Parity-time symmetry meets
  photonics: {A} new twist in non-{Hermitian} optics},\ }\href
  {https://doi.org/10.1209/0295-5075/120/64001} {\bibfield  {journal} {\bibinfo
   {journal} {Europhys. Lett.}\ }\textbf {\bibinfo {volume} {120}},\ \bibinfo
  {pages} {64001} (\bibinfo {year} {2018})}\BibitemShut {NoStop}%
\bibitem [{\citenamefont {Bender}\ and\ \citenamefont
  {Hook}(2024)}]{Bender2024}%
  \BibitemOpen
  \bibfield  {author} {\bibinfo {author} {\bibfnamefont {C.~M.}\ \bibnamefont
  {Bender}}\ and\ \bibinfo {author} {\bibfnamefont {D.~W.}\ \bibnamefont
  {Hook}},\ }\bibfield  {title} {\bibinfo {title} {$\mathcal{PT}$-symmetric
  quantum mechanics},\ }\href {https://doi.org/10.1103/RevModPhys.96.045002}
  {\bibfield  {journal} {\bibinfo  {journal} {Rev. Mod. Phys.}\ }\textbf
  {\bibinfo {volume} {96}},\ \bibinfo {pages} {045002} (\bibinfo {year}
  {2024})}\BibitemShut {NoStop}%
\bibitem [{\citenamefont {Chitambar}\ and\ \citenamefont
  {Gour}(2019)}]{chitambar2019quantum}%
  \BibitemOpen
  \bibfield  {author} {\bibinfo {author} {\bibfnamefont {E.}~\bibnamefont
  {Chitambar}}\ and\ \bibinfo {author} {\bibfnamefont {G.}~\bibnamefont
  {Gour}},\ }\bibfield  {title} {\bibinfo {title} {Quantum resource theories},\
  }\href {https://doi.org/10.1103/RevModPhys.91.025001} {\bibfield  {journal}
  {\bibinfo  {journal} {Rev. Mod. Phys.}\ }\textbf {\bibinfo {volume} {91}},\
  \bibinfo {pages} {025001} (\bibinfo {year} {2019})}\BibitemShut {NoStop}%
\bibitem [{\citenamefont {Veitch}\ \emph {et~al.}(2014)\citenamefont {Veitch},
  \citenamefont {Hamed~Mousavian}, \citenamefont {Gottesman},\ and\
  \citenamefont {Emerson}}]{veitch2014resource}%
  \BibitemOpen
  \bibfield  {author} {\bibinfo {author} {\bibfnamefont {V.}~\bibnamefont
  {Veitch}}, \bibinfo {author} {\bibfnamefont {S.~A.}\ \bibnamefont
  {Hamed~Mousavian}}, \bibinfo {author} {\bibfnamefont {D.}~\bibnamefont
  {Gottesman}},\ and\ \bibinfo {author} {\bibfnamefont {J.}~\bibnamefont
  {Emerson}},\ }\bibfield  {title} {\bibinfo {title} {The resource theory of
  stabilizer quantum computation},\ }\href
  {https://doi.org/10.1088/1367-2630/16/1/013009} {\bibfield  {journal}
  {\bibinfo  {journal} {New J. Phys.}\ }\textbf {\bibinfo {volume} {16}},\
  \bibinfo {pages} {013009} (\bibinfo {year} {2014})}\BibitemShut {NoStop}%
\bibitem [{\citenamefont {Howard}\ and\ \citenamefont
  {Campbell}(2017)}]{howard2017application}%
  \BibitemOpen
  \bibfield  {author} {\bibinfo {author} {\bibfnamefont {M.}~\bibnamefont
  {Howard}}\ and\ \bibinfo {author} {\bibfnamefont {E.}~\bibnamefont
  {Campbell}},\ }\bibfield  {title} {\bibinfo {title} {Application of a
  resource theory for magic states to fault-tolerant quantum computing},\
  }\href {https://doi.org/10.1103/PhysRevLett.118.090501} {\bibfield  {journal}
  {\bibinfo  {journal} {Phys. Rev. Lett.}\ }\textbf {\bibinfo {volume} {118}},\
  \bibinfo {pages} {090501} (\bibinfo {year} {2017})}\BibitemShut {NoStop}%
\bibitem [{\citenamefont {Liu}\ and\ \citenamefont
  {Winter}(2022)}]{liu2022many}%
  \BibitemOpen
  \bibfield  {author} {\bibinfo {author} {\bibfnamefont {Z.-W.}\ \bibnamefont
  {Liu}}\ and\ \bibinfo {author} {\bibfnamefont {A.}~\bibnamefont {Winter}},\
  }\bibfield  {title} {\bibinfo {title} {Many-body quantum magic},\ }\href
  {https://doi.org/10.1103/PRXQuantum.3.020333} {\bibfield  {journal} {\bibinfo
   {journal} {PRX Quantum}\ }\textbf {\bibinfo {volume} {3}},\ \bibinfo {pages}
  {020333} (\bibinfo {year} {2022})}\BibitemShut {NoStop}%
\bibitem [{\citenamefont {Leone}\ \emph {et~al.}(2022)\citenamefont {Leone},
  \citenamefont {Oliviero},\ and\ \citenamefont {Hamma}}]{leone2022stabilizer}%
  \BibitemOpen
  \bibfield  {author} {\bibinfo {author} {\bibfnamefont {L.}~\bibnamefont
  {Leone}}, \bibinfo {author} {\bibfnamefont {S.~F.~E.}\ \bibnamefont
  {Oliviero}},\ and\ \bibinfo {author} {\bibfnamefont {A.}~\bibnamefont
  {Hamma}},\ }\bibfield  {title} {\bibinfo {title} {Stabilizer {R\'e}nyi
  entropy},\ }\href {https://doi.org/10.1103/PhysRevLett.128.050402} {\bibfield
   {journal} {\bibinfo  {journal} {Phys. Rev. Lett.}\ }\textbf {\bibinfo
  {volume} {128}},\ \bibinfo {pages} {050402} (\bibinfo {year}
  {2022})}\BibitemShut {NoStop}%
\bibitem [{\citenamefont {Odavi\ifmmode~\acute{c}\else \'{c}\fi{}}\ \emph
  {et~al.}(2025)\citenamefont {Odavi\ifmmode~\acute{c}\else \'{c}\fi{}},
  \citenamefont {Viscardi},\ and\ \citenamefont
  {Hamma}}]{odavic2024stabilizer}%
  \BibitemOpen
  \bibfield  {author} {\bibinfo {author} {\bibfnamefont {J.}~\bibnamefont
  {Odavi\ifmmode~\acute{c}\else \'{c}\fi{}}}, \bibinfo {author} {\bibfnamefont
  {M.}~\bibnamefont {Viscardi}},\ and\ \bibinfo {author} {\bibfnamefont
  {A.}~\bibnamefont {Hamma}},\ }\bibfield  {title} {\bibinfo {title}
  {Stabilizer entropy in nonintegrable quantum evolutions},\ }\href
  {https://doi.org/10.1103/y9r6-dx7p} {\bibfield  {journal} {\bibinfo
  {journal} {Phys. Rev. B}\ }\textbf {\bibinfo {volume} {112}},\ \bibinfo
  {pages} {104301} (\bibinfo {year} {2025})}\BibitemShut {NoStop}%
\bibitem [{\citenamefont {Tarabunga}(2024)}]{tarabunga2024critical}%
  \BibitemOpen
  \bibfield  {author} {\bibinfo {author} {\bibfnamefont {P.~S.}\ \bibnamefont
  {Tarabunga}},\ }\bibfield  {title} {\bibinfo {title} {Critical behaviors of
  non-stabilizerness in quantum spin chains},\ }\href
  {https://doi.org/10.22331/q-2024-07-17-1413} {\bibfield  {journal} {\bibinfo
  {journal} {{Quantum}}\ }\textbf {\bibinfo {volume} {8}},\ \bibinfo {pages}
  {1413} (\bibinfo {year} {2024})}\BibitemShut {NoStop}%
\bibitem [{\citenamefont {Haug}\ and\ \citenamefont
  {Piroli}(2023)}]{haug2023stabilizer}%
  \BibitemOpen
  \bibfield  {author} {\bibinfo {author} {\bibfnamefont {T.}~\bibnamefont
  {Haug}}\ and\ \bibinfo {author} {\bibfnamefont {L.}~\bibnamefont {Piroli}},\
  }\bibfield  {title} {\bibinfo {title} {Stabilizer entropies and
  nonstabilizerness monotones},\ }\href
  {https://doi.org/10.22331/q-2023-08-28-1092} {\bibfield  {journal} {\bibinfo
  {journal} {{Quantum}}\ }\textbf {\bibinfo {volume} {7}},\ \bibinfo {pages}
  {1092} (\bibinfo {year} {2023})}\BibitemShut {NoStop}%
\bibitem [{\citenamefont {Sinibaldi}\ \emph {et~al.}(2025)\citenamefont
  {Sinibaldi}, \citenamefont {Mello}, \citenamefont {Collura},\ and\
  \citenamefont {Carleo}}]{sinibaldi2025non}%
  \BibitemOpen
  \bibfield  {author} {\bibinfo {author} {\bibfnamefont {A.}~\bibnamefont
  {Sinibaldi}}, \bibinfo {author} {\bibfnamefont {A.~F.}\ \bibnamefont
  {Mello}}, \bibinfo {author} {\bibfnamefont {M.}~\bibnamefont {Collura}},\
  and\ \bibinfo {author} {\bibfnamefont {G.}~\bibnamefont {Carleo}},\
  }\bibfield  {title} {\bibinfo {title} {Nonstabilizerness of neural quantum
  states},\ }\href {https://doi.org/10.1103/v5tw-yn1f} {\bibfield  {journal}
  {\bibinfo  {journal} {Phys. Rev. Res.}\ }\textbf {\bibinfo {volume} {7}},\
  \bibinfo {pages} {043289} (\bibinfo {year} {2025})}\BibitemShut {NoStop}%
\bibitem [{\citenamefont {Huang}\ \emph {et~al.}(2025)\citenamefont {Huang},
  \citenamefont {Qian},\ and\ \citenamefont {Qin}}]{huang2024non}%
  \BibitemOpen
  \bibfield  {author} {\bibinfo {author} {\bibfnamefont {J.}~\bibnamefont
  {Huang}}, \bibinfo {author} {\bibfnamefont {X.}~\bibnamefont {Qian}},\ and\
  \bibinfo {author} {\bibfnamefont {M.}~\bibnamefont {Qin}},\ }\bibfield
  {title} {\bibinfo {title} {Nonstabilizerness entanglement entropy: {A}
  measure of hardness in the classical simulation of quantum many-body systems
  with tensor network states},\ }\href {https://doi.org/10.1103/gxdn-zwrw}
  {\bibfield  {journal} {\bibinfo  {journal} {Phys. Rev. A}\ }\textbf {\bibinfo
  {volume} {112}},\ \bibinfo {pages} {012425} (\bibinfo {year}
  {2025})}\BibitemShut {NoStop}%
\bibitem [{\citenamefont {Gottesman}(1997)}]{gottesman1997stabilizer}%
  \BibitemOpen
  \bibfield  {author} {\bibinfo {author} {\bibfnamefont {D.}~\bibnamefont
  {Gottesman}},\ }\emph {\bibinfo {title} {Stabilizer Codes and Quantum Error
  Correction}},\ \href {https://doi.org/10.7907/rzr7-dt72} {Ph.D. thesis},\
  \bibinfo  {school} {California Institute of Technology} (\bibinfo {year}
  {1997}),\ \Eprint {https://arxiv.org/abs/quant-ph/9705052}
  {arXiv:quant-ph/9705052} \BibitemShut {NoStop}%
\bibitem [{\citenamefont {Aaronson}\ and\ \citenamefont
  {Gottesman}(2004)}]{Aaronson2004}%
  \BibitemOpen
  \bibfield  {author} {\bibinfo {author} {\bibfnamefont {S.}~\bibnamefont
  {Aaronson}}\ and\ \bibinfo {author} {\bibfnamefont {D.}~\bibnamefont
  {Gottesman}},\ }\bibfield  {title} {\bibinfo {title} {Improved simulation of
  stabilizer circuits},\ }\href {https://doi.org/10.1103/PhysRevA.70.052328}
  {\bibfield  {journal} {\bibinfo  {journal} {Phys. Rev. A}\ }\textbf {\bibinfo
  {volume} {70}},\ \bibinfo {pages} {052328} (\bibinfo {year}
  {2004})}\BibitemShut {NoStop}%
\bibitem [{\citenamefont {Dehaene}\ and\ \citenamefont
  {De~Moor}(2003)}]{dehaene2003clifford}%
  \BibitemOpen
  \bibfield  {author} {\bibinfo {author} {\bibfnamefont {J.}~\bibnamefont
  {Dehaene}}\ and\ \bibinfo {author} {\bibfnamefont {B.}~\bibnamefont
  {De~Moor}},\ }\bibfield  {title} {\bibinfo {title} {Clifford group,
  stabilizer states, and linear and quadratic operations over {GF(2)}},\ }\href
  {https://doi.org/10.1103/PhysRevA.68.042318} {\bibfield  {journal} {\bibinfo
  {journal} {Phys. Rev. A}\ }\textbf {\bibinfo {volume} {68}},\ \bibinfo
  {pages} {042318} (\bibinfo {year} {2003})}\BibitemShut {NoStop}%
\bibitem [{\citenamefont {Masot-Llima}\ and\ \citenamefont
  {Garcia-Saez}(2024)}]{masot2024stabilizer}%
  \BibitemOpen
  \bibfield  {author} {\bibinfo {author} {\bibfnamefont {S.}~\bibnamefont
  {Masot-Llima}}\ and\ \bibinfo {author} {\bibfnamefont {A.}~\bibnamefont
  {Garcia-Saez}},\ }\bibfield  {title} {\bibinfo {title} {Stabilizer tensor
  networks: {Universal} quantum simulator on a basis of stabilizer states},\
  }\href {https://doi.org/10.1103/PhysRevLett.133.230601} {\bibfield  {journal}
  {\bibinfo  {journal} {Phys. Rev. Lett.}\ }\textbf {\bibinfo {volume} {133}},\
  \bibinfo {pages} {230601} (\bibinfo {year} {2024})}\BibitemShut {NoStop}%
\bibitem [{\citenamefont {Brown}(2020)}]{brown2020fault}%
  \BibitemOpen
  \bibfield  {author} {\bibinfo {author} {\bibfnamefont {B.~J.}\ \bibnamefont
  {Brown}},\ }\bibfield  {title} {\bibinfo {title} {A fault-tolerant
  non-{Clifford} gate for the surface code in two dimensions},\ }\bibfield
  {journal} {\bibinfo  {journal} {Sci. Adv.}\ }\textbf {\bibinfo {volume}
  {6}},\ \href {https://doi.org/10.1126/sciadv.aay4929}
  {10.1126/sciadv.aay4929} (\bibinfo {year} {2020})\BibitemShut {NoStop}%
\bibitem [{\citenamefont {Oliviero}\ \emph {et~al.}(2022)\citenamefont
  {Oliviero}, \citenamefont {Leone},\ and\ \citenamefont
  {Hamma}}]{oliviero2022magic}%
  \BibitemOpen
  \bibfield  {author} {\bibinfo {author} {\bibfnamefont {S.~F.~E.}\
  \bibnamefont {Oliviero}}, \bibinfo {author} {\bibfnamefont {L.}~\bibnamefont
  {Leone}},\ and\ \bibinfo {author} {\bibfnamefont {A.}~\bibnamefont {Hamma}},\
  }\bibfield  {title} {\bibinfo {title} {Magic-state resource theory for the
  ground state of the transverse-field {Ising} model},\ }\href
  {https://doi.org/10.1103/PhysRevA.106.042426} {\bibfield  {journal} {\bibinfo
   {journal} {Phys. Rev. A}\ }\textbf {\bibinfo {volume} {106}},\ \bibinfo
  {pages} {042426} (\bibinfo {year} {2022})}\BibitemShut {NoStop}%
\bibitem [{\citenamefont {Odavić}\ \emph {et~al.}(2023)\citenamefont
  {Odavić}, \citenamefont {Haug}, \citenamefont {Torre}, \citenamefont
  {Hamma}, \citenamefont {Franchini},\ and\ \citenamefont
  {Giampaolo}}]{odavic2023complexity}%
  \BibitemOpen
  \bibfield  {author} {\bibinfo {author} {\bibfnamefont {J.}~\bibnamefont
  {Odavić}}, \bibinfo {author} {\bibfnamefont {T.}~\bibnamefont {Haug}},
  \bibinfo {author} {\bibfnamefont {G.}~\bibnamefont {Torre}}, \bibinfo
  {author} {\bibfnamefont {A.}~\bibnamefont {Hamma}}, \bibinfo {author}
  {\bibfnamefont {F.}~\bibnamefont {Franchini}},\ and\ \bibinfo {author}
  {\bibfnamefont {S.~M.}\ \bibnamefont {Giampaolo}},\ }\bibfield  {title}
  {\bibinfo {title} {{Complexity of frustration: {A} new source of non-local
  non-stabilizerness}},\ }\href {https://doi.org/10.21468/SciPostPhys.15.4.131}
  {\bibfield  {journal} {\bibinfo  {journal} {SciPost Phys.}\ }\textbf
  {\bibinfo {volume} {15}},\ \bibinfo {pages} {131} (\bibinfo {year}
  {2023})}\BibitemShut {NoStop}%
\bibitem [{\citenamefont {Sticlet}\ \emph {et~al.}(2025)\citenamefont
  {Sticlet}, \citenamefont {D\'ora}, \citenamefont {Szombathy}, \citenamefont
  {Zar\'and},\ and\ \citenamefont {Moca}}]{sticlet2025non}%
  \BibitemOpen
  \bibfield  {author} {\bibinfo {author} {\bibfnamefont {D.}~\bibnamefont
  {Sticlet}}, \bibinfo {author} {\bibfnamefont {B.}~\bibnamefont {D\'ora}},
  \bibinfo {author} {\bibfnamefont {D.}~\bibnamefont {Szombathy}}, \bibinfo
  {author} {\bibfnamefont {G.}~\bibnamefont {Zar\'and}},\ and\ \bibinfo
  {author} {\bibfnamefont {C.~P.}\ \bibnamefont {Moca}},\ }\bibfield  {title}
  {\bibinfo {title} {Nonstabilizerness in open {XXZ} spin chains: Universal
  scaling and dynamics},\ }\href {https://doi.org/10.1103/96bk-xf8p} {\bibfield
   {journal} {\bibinfo  {journal} {Phys. Rev. Res.}\ }\textbf {\bibinfo
  {volume} {7}},\ \bibinfo {pages} {043130} (\bibinfo {year}
  {2025})}\BibitemShut {NoStop}%
\bibitem [{\citenamefont {Sarkis}\ and\ \citenamefont
  {Tkatchenko}(2025)}]{sarkis2025molecules}%
  \BibitemOpen
  \bibfield  {author} {\bibinfo {author} {\bibfnamefont {M.}~\bibnamefont
  {Sarkis}}\ and\ \bibinfo {author} {\bibfnamefont {A.}~\bibnamefont
  {Tkatchenko}},\ }\href {https://doi.org/10.48550/arxiv.2504.06673} {\bibinfo
  {title} {Are molecules magical? {Non}-stabilizerness in molecular bonding}}
  (\bibinfo {year} {2025}),\ \Eprint {https://arxiv.org/abs/2504.06673}
  {arXiv:2504.06673 [quant-ph]} \BibitemShut {NoStop}%
\bibitem [{\citenamefont {Moca}\ \emph {et~al.}(2026)\citenamefont {Moca},
  \citenamefont {Sticlet}, \citenamefont {Dóra}, \citenamefont {Valli},
  \citenamefont {Szombathy},\ and\ \citenamefont {Zaránd}}]{moca2025non}%
  \BibitemOpen
  \bibfield  {author} {\bibinfo {author} {\bibfnamefont {C.~P.}\ \bibnamefont
  {Moca}}, \bibinfo {author} {\bibfnamefont {D.}~\bibnamefont {Sticlet}},
  \bibinfo {author} {\bibfnamefont {B.}~\bibnamefont {Dóra}}, \bibinfo
  {author} {\bibfnamefont {A.}~\bibnamefont {Valli}}, \bibinfo {author}
  {\bibfnamefont {D.}~\bibnamefont {Szombathy}},\ and\ \bibinfo {author}
  {\bibfnamefont {G.}~\bibnamefont {Zaránd}},\ }\bibfield  {title} {\bibinfo
  {title} {Nonstabilizerness generation in a multiparticle quantum walk},\
  }\href {https://doi.org/10.1103/nzrp-49mr} {\bibfield  {journal} {\bibinfo
  {journal} {Phys. Rev. B}\ }\textbf {\bibinfo {volume} {113}},\ \bibinfo
  {pages} {075142} (\bibinfo {year} {2026})}\BibitemShut {NoStop}%
\bibitem [{\citenamefont {Iannotti}\ \emph {et~al.}(2025)\citenamefont
  {Iannotti}, \citenamefont {Esposito}, \citenamefont {Campos~Venuti},\ and\
  \citenamefont {Hamma}}]{Iannotti2025}%
  \BibitemOpen
  \bibfield  {author} {\bibinfo {author} {\bibfnamefont {D.}~\bibnamefont
  {Iannotti}}, \bibinfo {author} {\bibfnamefont {G.}~\bibnamefont {Esposito}},
  \bibinfo {author} {\bibfnamefont {L.}~\bibnamefont {Campos~Venuti}},\ and\
  \bibinfo {author} {\bibfnamefont {A.}~\bibnamefont {Hamma}},\ }\bibfield
  {title} {\bibinfo {title} {Entanglement and {S}tabilizer entropies of random
  bipartite pure quantum states},\ }\href
  {https://doi.org/10.22331/q-2025-07-21-1797} {\bibfield  {journal} {\bibinfo
  {journal} {{Quantum}}\ }\textbf {\bibinfo {volume} {9}},\ \bibinfo {pages}
  {1797} (\bibinfo {year} {2025})}\BibitemShut {NoStop}%
\bibitem [{\citenamefont {Turkeshi}\ \emph
  {et~al.}(2025{\natexlab{a}})\citenamefont {Turkeshi}, \citenamefont
  {Dymarsky},\ and\ \citenamefont {Sierant}}]{Turkeshi2025a}%
  \BibitemOpen
  \bibfield  {author} {\bibinfo {author} {\bibfnamefont {X.}~\bibnamefont
  {Turkeshi}}, \bibinfo {author} {\bibfnamefont {A.}~\bibnamefont {Dymarsky}},\
  and\ \bibinfo {author} {\bibfnamefont {P.}~\bibnamefont {Sierant}},\
  }\bibfield  {title} {\bibinfo {title} {Pauli spectrum and nonstabilizerness
  of typical quantum many-body states},\ }\href
  {https://doi.org/10.1103/PhysRevB.111.054301} {\bibfield  {journal} {\bibinfo
   {journal} {Phys. Rev. B}\ }\textbf {\bibinfo {volume} {111}},\ \bibinfo
  {pages} {054301} (\bibinfo {year} {2025}{\natexlab{a}})}\BibitemShut
  {NoStop}%
\bibitem [{\citenamefont {Bera}\ and\ \citenamefont
  {Schirò}(2025)}]{bera2025non}%
  \BibitemOpen
  \bibfield  {author} {\bibinfo {author} {\bibfnamefont {S.}~\bibnamefont
  {Bera}}\ and\ \bibinfo {author} {\bibfnamefont {M.}~\bibnamefont {Schirò}},\
  }\bibfield  {title} {\bibinfo {title} {{Non-stabilizerness of
  Sachdev-Ye-Kitaev model}},\ }\href
  {https://doi.org/10.21468/SciPostPhys.19.6.159} {\bibfield  {journal}
  {\bibinfo  {journal} {SciPost Phys.}\ }\textbf {\bibinfo {volume} {19}},\
  \bibinfo {pages} {159} (\bibinfo {year} {2025})}\BibitemShut {NoStop}%
\bibitem [{\citenamefont {Viscardi}\ \emph {et~al.}(2026)\citenamefont
  {Viscardi}, \citenamefont {Dalmonte}, \citenamefont {Hamma},\ and\
  \citenamefont {Tirrito}}]{Viscardi2025}%
  \BibitemOpen
  \bibfield  {author} {\bibinfo {author} {\bibfnamefont {M.}~\bibnamefont
  {Viscardi}}, \bibinfo {author} {\bibfnamefont {M.}~\bibnamefont {Dalmonte}},
  \bibinfo {author} {\bibfnamefont {A.}~\bibnamefont {Hamma}},\ and\ \bibinfo
  {author} {\bibfnamefont {E.}~\bibnamefont {Tirrito}},\ }\bibfield  {title}
  {\bibinfo {title} {{Interplay of entanglement structures and stabilizer
  entropy in spin models}},\ }\href
  {https://doi.org/10.21468/SciPostPhysCore.9.1.012} {\bibfield  {journal}
  {\bibinfo  {journal} {SciPost Phys. Core}\ }\textbf {\bibinfo {volume} {9}},\
  \bibinfo {pages} {012} (\bibinfo {year} {2026})}\BibitemShut {NoStop}%
\bibitem [{\citenamefont {Catalano}\ \emph {et~al.}(2025)\citenamefont
  {Catalano}, \citenamefont {Odavić}, \citenamefont {Torre}, \citenamefont
  {Hamma}, \citenamefont {Franchini},\ and\ \citenamefont
  {Giampaolo}}]{Catalano2025}%
  \BibitemOpen
  \bibfield  {author} {\bibinfo {author} {\bibfnamefont {A.~G.}\ \bibnamefont
  {Catalano}}, \bibinfo {author} {\bibfnamefont {J.}~\bibnamefont {Odavić}},
  \bibinfo {author} {\bibfnamefont {G.}~\bibnamefont {Torre}}, \bibinfo
  {author} {\bibfnamefont {A.}~\bibnamefont {Hamma}}, \bibinfo {author}
  {\bibfnamefont {F.}~\bibnamefont {Franchini}},\ and\ \bibinfo {author}
  {\bibfnamefont {S.~M.}\ \bibnamefont {Giampaolo}},\ }\bibfield  {title}
  {\bibinfo {title} {Magic phase transition and non-local complexity in
  generalized {W} {State}},\ }\href
  {https://doi.org/10.21468/SciPostPhysCore.8.4.078} {\bibfield  {journal}
  {\bibinfo  {journal} {SciPost Phys. Core}\ }\textbf {\bibinfo {volume} {8}},\
  \bibinfo {pages} {078} (\bibinfo {year} {2025})}\BibitemShut {NoStop}%
\bibitem [{\citenamefont {Heiss}(2004)}]{heiss2004exceptional}%
  \BibitemOpen
  \bibfield  {author} {\bibinfo {author} {\bibfnamefont {W.~D.}\ \bibnamefont
  {Heiss}},\ }\bibfield  {title} {\bibinfo {title} {Exceptional points of
  non-{Hermitian} operators},\ }\href
  {https://doi.org/10.1088/0305-4470/37/6/034} {\bibfield  {journal} {\bibinfo
  {journal} {J. Phys. A: Math. Gen.}\ }\textbf {\bibinfo {volume} {37}},\
  \bibinfo {pages} {2455–2464} (\bibinfo {year} {2004})}\BibitemShut
  {NoStop}%
\bibitem [{\citenamefont {Berry}(2004)}]{Berry2004}%
  \BibitemOpen
  \bibfield  {author} {\bibinfo {author} {\bibfnamefont {M.~V.}\ \bibnamefont
  {Berry}},\ }\bibfield  {title} {\bibinfo {title} {Physics of nonhermitian
  degeneracies},\ }\href {https://doi.org/10.1023/B:CJOP.0000044002.05657.04}
  {\bibfield  {journal} {\bibinfo  {journal} {Czech. J. Phys.}\ }\textbf
  {\bibinfo {volume} {54}},\ \bibinfo {pages} {1039} (\bibinfo {year}
  {2004})}\BibitemShut {NoStop}%
\bibitem [{\citenamefont {Minganti}\ \emph {et~al.}(2019)\citenamefont
  {Minganti}, \citenamefont {Miranowicz}, \citenamefont {Chhajlany},\ and\
  \citenamefont {Nori}}]{minganti2019quantum}%
  \BibitemOpen
  \bibfield  {author} {\bibinfo {author} {\bibfnamefont {F.}~\bibnamefont
  {Minganti}}, \bibinfo {author} {\bibfnamefont {A.}~\bibnamefont
  {Miranowicz}}, \bibinfo {author} {\bibfnamefont {R.~W.}\ \bibnamefont
  {Chhajlany}},\ and\ \bibinfo {author} {\bibfnamefont {F.}~\bibnamefont
  {Nori}},\ }\bibfield  {title} {\bibinfo {title} {Quantum exceptional points
  of non-{Hermitian} {Hamiltonians} and {Liouvillians}: {The} effects of
  quantum jumps},\ }\href {https://doi.org/10.1103/PhysRevA.100.062131}
  {\bibfield  {journal} {\bibinfo  {journal} {Phys. Rev. A}\ }\textbf {\bibinfo
  {volume} {100}},\ \bibinfo {pages} {062131} (\bibinfo {year}
  {2019})}\BibitemShut {NoStop}%
\bibitem [{\citenamefont {Ding}\ \emph {et~al.}(2022)\citenamefont {Ding},
  \citenamefont {Fang},\ and\ \citenamefont {Ma}}]{ding2022non}%
  \BibitemOpen
  \bibfield  {author} {\bibinfo {author} {\bibfnamefont {K.}~\bibnamefont
  {Ding}}, \bibinfo {author} {\bibfnamefont {C.}~\bibnamefont {Fang}},\ and\
  \bibinfo {author} {\bibfnamefont {G.}~\bibnamefont {Ma}},\ }\bibfield
  {title} {\bibinfo {title} {Non-{Hermitian} topology and exceptional-point
  geometries},\ }\href {https://doi.org/10.1038/s42254-022-00516-5} {\bibfield
  {journal} {\bibinfo  {journal} {Nat. Rev. Phys.}\ }\textbf {\bibinfo {volume}
  {4}},\ \bibinfo {pages} {745–760} (\bibinfo {year} {2022})}\BibitemShut
  {NoStop}%
\bibitem [{\citenamefont {Starkov}\ \emph {et~al.}(2023)\citenamefont
  {Starkov}, \citenamefont {Fistul},\ and\ \citenamefont
  {Eremin}}]{starkov2023quantum}%
  \BibitemOpen
  \bibfield  {author} {\bibinfo {author} {\bibfnamefont {G.~A.}\ \bibnamefont
  {Starkov}}, \bibinfo {author} {\bibfnamefont {M.~V.}\ \bibnamefont
  {Fistul}},\ and\ \bibinfo {author} {\bibfnamefont {I.~M.}\ \bibnamefont
  {Eremin}},\ }\bibfield  {title} {\bibinfo {title} {Quantum phase transitions
  in non-{Hermitian} {PT}-symmetric transverse-field {Ising} spin chains},\
  }\href {https://doi.org/10.1016/j.aop.2023.169268} {\bibfield  {journal}
  {\bibinfo  {journal} {Ann. Phys. (N. Y.)}\ }\textbf {\bibinfo {volume}
  {456}},\ \bibinfo {pages} {169268} (\bibinfo {year} {2023})}\BibitemShut
  {NoStop}%
\bibitem [{\citenamefont {Lu}\ \emph {et~al.}(2024)\citenamefont {Lu},
  \citenamefont {Deng}, \citenamefont {Kou},\ and\ \citenamefont
  {Sun}}]{lu2024many}%
  \BibitemOpen
  \bibfield  {author} {\bibinfo {author} {\bibfnamefont {C.-Z.}\ \bibnamefont
  {Lu}}, \bibinfo {author} {\bibfnamefont {X.}~\bibnamefont {Deng}}, \bibinfo
  {author} {\bibfnamefont {S.-P.}\ \bibnamefont {Kou}},\ and\ \bibinfo {author}
  {\bibfnamefont {G.}~\bibnamefont {Sun}},\ }\bibfield  {title} {\bibinfo
  {title} {Many-body phase transitions in a non-{Hermitian} {Ising} chain},\
  }\href {https://doi.org/10.1103/PhysRevB.110.014441} {\bibfield  {journal}
  {\bibinfo  {journal} {Phys. Rev. B}\ }\textbf {\bibinfo {volume} {110}},\
  \bibinfo {pages} {014441} (\bibinfo {year} {2024})}\BibitemShut {NoStop}%
\bibitem [{\citenamefont {Tirrito}\ \emph
  {et~al.}(2025{\natexlab{a}})\citenamefont {Tirrito}, \citenamefont {Lumia},
  \citenamefont {Paviglianiti}, \citenamefont {Lami}, \citenamefont {Silva},
  \citenamefont {Turkeshi},\ and\ \citenamefont {Collura}}]{Tirrito2025}%
  \BibitemOpen
  \bibfield  {author} {\bibinfo {author} {\bibfnamefont {E.}~\bibnamefont
  {Tirrito}}, \bibinfo {author} {\bibfnamefont {L.}~\bibnamefont {Lumia}},
  \bibinfo {author} {\bibfnamefont {A.}~\bibnamefont {Paviglianiti}}, \bibinfo
  {author} {\bibfnamefont {G.}~\bibnamefont {Lami}}, \bibinfo {author}
  {\bibfnamefont {A.}~\bibnamefont {Silva}}, \bibinfo {author} {\bibfnamefont
  {X.}~\bibnamefont {Turkeshi}},\ and\ \bibinfo {author} {\bibfnamefont
  {M.}~\bibnamefont {Collura}},\ }\href {https://arxiv.org/abs/2507.07179}
  {\bibinfo {title} {Magic phase transitions in monitored gaussian fermions}}
  (\bibinfo {year} {2025}{\natexlab{a}}),\ \Eprint
  {https://arxiv.org/abs/2507.07179} {arXiv:2507.07179 [quant-ph]} \BibitemShut
  {NoStop}%
\bibitem [{\citenamefont {Ashida}\ \emph {et~al.}(2017)\citenamefont {Ashida},
  \citenamefont {Furukawa},\ and\ \citenamefont {Ueda}}]{Ashida2017}%
  \BibitemOpen
  \bibfield  {author} {\bibinfo {author} {\bibfnamefont {Y.}~\bibnamefont
  {Ashida}}, \bibinfo {author} {\bibfnamefont {S.}~\bibnamefont {Furukawa}},\
  and\ \bibinfo {author} {\bibfnamefont {M.}~\bibnamefont {Ueda}},\ }\bibfield
  {title} {\bibinfo {title} {Parity-time-symmetric quantum critical
  phenomena},\ }\bibfield  {journal} {\bibinfo  {journal} {Nat. Commun.}\
  }\textbf {\bibinfo {volume} {8}},\ \href
  {https://doi.org/10.1038/ncomms15791} {10.1038/ncomms15791} (\bibinfo {year}
  {2017})\BibitemShut {NoStop}%
\bibitem [{\citenamefont {D\'ora}\ \emph {et~al.}(2022)\citenamefont {D\'ora},
  \citenamefont {Sticlet},\ and\ \citenamefont {Moca}}]{dora2022correlations}%
  \BibitemOpen
  \bibfield  {author} {\bibinfo {author} {\bibfnamefont {B.}~\bibnamefont
  {D\'ora}}, \bibinfo {author} {\bibfnamefont {D.}~\bibnamefont {Sticlet}},\
  and\ \bibinfo {author} {\bibfnamefont {C.~P.}\ \bibnamefont {Moca}},\
  }\bibfield  {title} {\bibinfo {title} {Correlations at {PT}-symmetric quantum
  critical point},\ }\href {https://doi.org/10.1103/PhysRevLett.128.146804}
  {\bibfield  {journal} {\bibinfo  {journal} {Phys. Rev. Lett.}\ }\textbf
  {\bibinfo {volume} {128}},\ \bibinfo {pages} {146804} (\bibinfo {year}
  {2022})}\BibitemShut {NoStop}%
\bibitem [{\citenamefont {Fortin}\ \emph {et~al.}(2016)\citenamefont {Fortin},
  \citenamefont {Holik},\ and\ \citenamefont {Vanni}}]{fortin2016non}%
  \BibitemOpen
  \bibfield  {author} {\bibinfo {author} {\bibfnamefont {S.}~\bibnamefont
  {Fortin}}, \bibinfo {author} {\bibfnamefont {F.}~\bibnamefont {Holik}},\ and\
  \bibinfo {author} {\bibfnamefont {L.}~\bibnamefont {Vanni}},\ }\bibfield
  {title} {\bibinfo {title} {Non-unitary evolution of quantum logics},\ }in\
  \href {https://doi.org/10.1007/978-3-319-31356-6_14} {\emph {\bibinfo
  {booktitle} {Non-Hermitian Hamiltonians in Quantum Physics}}},\ \bibinfo
  {series} {Springer Proceedings in Physics}, Vol.\ \bibinfo {volume} {184},\
  \bibinfo {editor} {edited by\ \bibinfo {editor} {\bibfnamefont
  {F.}~\bibnamefont {Bagarello}}, \bibinfo {editor} {\bibfnamefont
  {R.}~\bibnamefont {Passante}},\ and\ \bibinfo {editor} {\bibfnamefont
  {C.}~\bibnamefont {Trapani}}}\ (\bibinfo  {publisher} {Springer},\ \bibinfo
  {address} {Cham},\ \bibinfo {year} {2016})\ pp.\ \bibinfo {pages}
  {219--234}\BibitemShut {NoStop}%
\bibitem [{\citenamefont {Williams}(2004)}]{williams2004probabilistic}%
  \BibitemOpen
  \bibfield  {author} {\bibinfo {author} {\bibfnamefont {C.~P.}\ \bibnamefont
  {Williams}},\ }\bibfield  {title} {\bibinfo {title} {Probabilistic nonunitary
  quantum computing},\ }in\ \href {https://doi.org/10.1117/12.542413} {\emph
  {\bibinfo {booktitle} {Quantum Information and Computation II}}},\ \bibinfo
  {series} {Proc. SPIE}, Vol.\ \bibinfo {volume} {5436},\ \bibinfo {editor}
  {edited by\ \bibinfo {editor} {\bibfnamefont {E.}~\bibnamefont {Donkor}},
  \bibinfo {editor} {\bibfnamefont {A.~R.}\ \bibnamefont {Pirich}},\ and\
  \bibinfo {editor} {\bibfnamefont {H.~E.}\ \bibnamefont {Brandt}}}\ (\bibinfo
  {publisher} {SPIE},\ \bibinfo {address} {Bellingham, WA},\ \bibinfo {year}
  {2004})\ pp.\ \bibinfo {pages} {297--306}\BibitemShut {NoStop}%
\bibitem [{\citenamefont {Turkeshi}\ and\ \citenamefont
  {Schir\`o}(2023)}]{turkeshi2023entanglement}%
  \BibitemOpen
  \bibfield  {author} {\bibinfo {author} {\bibfnamefont {X.}~\bibnamefont
  {Turkeshi}}\ and\ \bibinfo {author} {\bibfnamefont {M.}~\bibnamefont
  {Schir\`o}},\ }\bibfield  {title} {\bibinfo {title} {Entanglement and
  correlation spreading in non-{Hermitian} spin chains},\ }\href
  {https://doi.org/10.1103/PhysRevB.107.L020403} {\bibfield  {journal}
  {\bibinfo  {journal} {Phys. Rev. B}\ }\textbf {\bibinfo {volume} {107}},\
  \bibinfo {pages} {L020403} (\bibinfo {year} {2023})}\BibitemShut {NoStop}%
\bibitem [{\citenamefont {Meden}\ \emph {et~al.}(2023)\citenamefont {Meden},
  \citenamefont {Grunwald},\ and\ \citenamefont {Kennes}}]{Meden2023}%
  \BibitemOpen
  \bibfield  {author} {\bibinfo {author} {\bibfnamefont {V.}~\bibnamefont
  {Meden}}, \bibinfo {author} {\bibfnamefont {L.}~\bibnamefont {Grunwald}},\
  and\ \bibinfo {author} {\bibfnamefont {D.~M.}\ \bibnamefont {Kennes}},\
  }\bibfield  {title} {\bibinfo {title}
  {{$\mathcal{P}$}{$\mathcal{T}$}-symmetric, non-{Hermitian} quantum many-body
  physics—a methodological perspective},\ }\href
  {https://doi.org/10.1088/1361-6633/ad05f3} {\bibfield  {journal} {\bibinfo
  {journal} {Reports on Progress in Physics}\ }\textbf {\bibinfo {volume}
  {86}},\ \bibinfo {pages} {124501} (\bibinfo {year} {2023})}\BibitemShut
  {NoStop}%
\bibitem [{\citenamefont {Dalibard}\ \emph {et~al.}(1992)\citenamefont
  {Dalibard}, \citenamefont {Castin},\ and\ \citenamefont
  {M\o{}lmer}}]{Dalibard1992}%
  \BibitemOpen
  \bibfield  {author} {\bibinfo {author} {\bibfnamefont {J.}~\bibnamefont
  {Dalibard}}, \bibinfo {author} {\bibfnamefont {Y.}~\bibnamefont {Castin}},\
  and\ \bibinfo {author} {\bibfnamefont {K.}~\bibnamefont {M\o{}lmer}},\
  }\bibfield  {title} {\bibinfo {title} {Wave-function approach to dissipative
  processes in quantum optics},\ }\href
  {https://doi.org/10.1103/PhysRevLett.68.580} {\bibfield  {journal} {\bibinfo
  {journal} {Phys. Rev. Lett.}\ }\textbf {\bibinfo {volume} {68}},\ \bibinfo
  {pages} {580} (\bibinfo {year} {1992})}\BibitemShut {NoStop}%
\bibitem [{\citenamefont {Carmichael}(1993)}]{Carmichael1993}%
  \BibitemOpen
  \bibfield  {author} {\bibinfo {author} {\bibfnamefont {H.}~\bibnamefont
  {Carmichael}},\ }\href {https://doi.org/10.1007/978-3-540-47620-7} {\emph
  {\bibinfo {title} {An Open Systems Approach to Quantum Optics}}},\ \bibinfo
  {series} {Lecture Notes in Physics Monographs}, Vol.~\bibinfo {volume} {18}\
  (\bibinfo  {publisher} {Springer-Verlag},\ \bibinfo {address} {Berlin,
  Heidelberg},\ \bibinfo {year} {1993})\BibitemShut {NoStop}%
\bibitem [{\citenamefont {White}(1992)}]{White1992}%
  \BibitemOpen
  \bibfield  {author} {\bibinfo {author} {\bibfnamefont {S.~R.}\ \bibnamefont
  {White}},\ }\bibfield  {title} {\bibinfo {title} {Density matrix formulation
  for quantum renormalization groups},\ }\href
  {https://doi.org/10.1103/PhysRevLett.69.2863} {\bibfield  {journal} {\bibinfo
   {journal} {Phys. Rev. Lett.}\ }\textbf {\bibinfo {volume} {69}},\ \bibinfo
  {pages} {2863} (\bibinfo {year} {1992})}\BibitemShut {NoStop}%
\bibitem [{\citenamefont {Fishman}\ \emph {et~al.}(2022)\citenamefont
  {Fishman}, \citenamefont {White},\ and\ \citenamefont
  {Stoudenmire}}]{fishman2022itensor}%
  \BibitemOpen
  \bibfield  {author} {\bibinfo {author} {\bibfnamefont {M.}~\bibnamefont
  {Fishman}}, \bibinfo {author} {\bibfnamefont {S.~R.}\ \bibnamefont {White}},\
  and\ \bibinfo {author} {\bibfnamefont {E.~M.}\ \bibnamefont {Stoudenmire}},\
  }\bibfield  {title} {\bibinfo {title} {The {ITensor} software library for
  tensor network calculations},\ }\href
  {https://doi.org/10.21468/SciPostPhysCodeb.4} {\bibfield  {journal} {\bibinfo
   {journal} {SciPost Phys. Codebases}\ ,\ \bibinfo {pages} {4}} (\bibinfo
  {year} {2022})}\BibitemShut {NoStop}%
\bibitem [{\citenamefont {Schollwöck}(2011)}]{schollwock2011density}%
  \BibitemOpen
  \bibfield  {author} {\bibinfo {author} {\bibfnamefont {U.}~\bibnamefont
  {Schollwöck}},\ }\bibfield  {title} {\bibinfo {title} {The density-matrix
  renormalization group in the age of matrix product states},\ }\href
  {https://doi.org/10.1016/j.aop.2010.09.012} {\bibfield  {journal} {\bibinfo
  {journal} {Ann. Phys (N. Y.)}\ }\textbf {\bibinfo {volume} {326}},\ \bibinfo
  {pages} {96} (\bibinfo {year} {2011})}\BibitemShut {NoStop}%
\bibitem [{\citenamefont {Chib}\ and\ \citenamefont
  {Greenberg}(1995)}]{chib1995understanding}%
  \BibitemOpen
  \bibfield  {author} {\bibinfo {author} {\bibfnamefont {S.}~\bibnamefont
  {Chib}}\ and\ \bibinfo {author} {\bibfnamefont {E.}~\bibnamefont
  {Greenberg}},\ }\bibfield  {title} {\bibinfo {title} {Understanding the
  {M}etropolis-{H}astings algorithm},\ }\href
  {https://doi.org/10.1080/00031305.1995.10476177} {\bibfield  {journal}
  {\bibinfo  {journal} {Amer. Statist.}\ }\textbf {\bibinfo {volume} {49}},\
  \bibinfo {pages} {327} (\bibinfo {year} {1995})}\BibitemShut {NoStop}%
\bibitem [{\citenamefont {Robert}\ and\ \citenamefont
  {Casella}(2009)}]{robert2009metropolis}%
  \BibitemOpen
  \bibfield  {author} {\bibinfo {author} {\bibfnamefont {C.~P.}\ \bibnamefont
  {Robert}}\ and\ \bibinfo {author} {\bibfnamefont {G.}~\bibnamefont
  {Casella}},\ }\bibfield  {title} {\bibinfo {title} {Metropolis-{Hastings}
  algorithms},\ }in\ \href {https://doi.org/10.1007/978-1-4419-1576-4_6} {\emph
  {\bibinfo {booktitle} {Introducing {Monte} {Carlo} Methods with {R}}}}\
  (\bibinfo  {publisher} {Springer},\ \bibinfo {address} {New York},\ \bibinfo
  {year} {2009})\ pp.\ \bibinfo {pages} {167--197}\BibitemShut {NoStop}%
\bibitem [{\citenamefont {Roberts}\ and\ \citenamefont
  {Rosenthal}(2001)}]{roberts2001optimal}%
  \BibitemOpen
  \bibfield  {author} {\bibinfo {author} {\bibfnamefont {G.~O.}\ \bibnamefont
  {Roberts}}\ and\ \bibinfo {author} {\bibfnamefont {J.~S.}\ \bibnamefont
  {Rosenthal}},\ }\bibfield  {title} {\bibinfo {title} {Optimal scaling for
  various {M}etropolis-{H}astings algorithms},\ }\href
  {https://doi.org/10.1214/ss/1015346320} {\bibfield  {journal} {\bibinfo
  {journal} {Statist. Sci.}\ }\textbf {\bibinfo {volume} {16}},\ \bibinfo
  {pages} {351} (\bibinfo {year} {2001})}\BibitemShut {NoStop}%
\bibitem [{\citenamefont {Tarabunga}\ \emph {et~al.}(2023)\citenamefont
  {Tarabunga}, \citenamefont {Tirrito}, \citenamefont {Chanda},\ and\
  \citenamefont {Dalmonte}}]{Tarabunga2023}%
  \BibitemOpen
  \bibfield  {author} {\bibinfo {author} {\bibfnamefont {P.~S.}\ \bibnamefont
  {Tarabunga}}, \bibinfo {author} {\bibfnamefont {E.}~\bibnamefont {Tirrito}},
  \bibinfo {author} {\bibfnamefont {T.}~\bibnamefont {Chanda}},\ and\ \bibinfo
  {author} {\bibfnamefont {M.}~\bibnamefont {Dalmonte}},\ }\bibfield  {title}
  {\bibinfo {title} {Many-body magic via {P}auli-{M}arkov chains---{F}rom
  criticality to gauge theories},\ }\href
  {https://doi.org/10.1103/PRXQuantum.4.040317} {\bibfield  {journal} {\bibinfo
   {journal} {PRX Quantum}\ }\textbf {\bibinfo {volume} {4}},\ \bibinfo {pages}
  {040317} (\bibinfo {year} {2023})}\BibitemShut {NoStop}%
\bibitem [{\citenamefont {Pfeuty}(1970)}]{pfeuty1970one}%
  \BibitemOpen
  \bibfield  {author} {\bibinfo {author} {\bibfnamefont {P.}~\bibnamefont
  {Pfeuty}},\ }\bibfield  {title} {\bibinfo {title} {The one-dimensional
  {Ising} model with a transverse field},\ }\href
  {https://doi.org/10.1016/0003-4916(70)90270-8} {\bibfield  {journal}
  {\bibinfo  {journal} {Ann. Phys-New. York.}\ }\textbf {\bibinfo {volume}
  {57}},\ \bibinfo {pages} {79} (\bibinfo {year} {1970})}\BibitemShut {NoStop}%
\bibitem [{\citenamefont {Yang}\ \emph {et~al.}(2022)\citenamefont {Yang},
  \citenamefont {Wang}, \citenamefont {Yang}, \citenamefont {Guo},
  \citenamefont {Wang}, \citenamefont {Sun},\ and\ \citenamefont
  {Kou}}]{yang2022hidden}%
  \BibitemOpen
  \bibfield  {author} {\bibinfo {author} {\bibfnamefont {F.}~\bibnamefont
  {Yang}}, \bibinfo {author} {\bibfnamefont {H.}~\bibnamefont {Wang}}, \bibinfo
  {author} {\bibfnamefont {M.-L.}\ \bibnamefont {Yang}}, \bibinfo {author}
  {\bibfnamefont {C.-X.}\ \bibnamefont {Guo}}, \bibinfo {author} {\bibfnamefont
  {X.-R.}\ \bibnamefont {Wang}}, \bibinfo {author} {\bibfnamefont {G.-Y.}\
  \bibnamefont {Sun}},\ and\ \bibinfo {author} {\bibfnamefont {S.-P.}\
  \bibnamefont {Kou}},\ }\bibfield  {title} {\bibinfo {title} {Hidden
  continuous quantum phase transition without gap closing in non-{Hermitian}
  transverse {Ising} model},\ }\href {https://doi.org/10.1088/1367-2630/ac652f}
  {\bibfield  {journal} {\bibinfo  {journal} {New J. Phys.}\ }\textbf {\bibinfo
  {volume} {24}},\ \bibinfo {pages} {043046} (\bibinfo {year}
  {2022})}\BibitemShut {NoStop}%
\bibitem [{Note1()}]{Note1}%
  \BibitemOpen
  \bibinfo {note} {An analysis of the excited states reveals the presence of
  the third critical line, in the $\protect \mathcal {PT}$-broken regime
  ($\gamma >1$), where the first and second excited states coalesce and acquire
  complex energies, again through second-order exceptional points.}\BibitemShut
  {Stop}%
\bibitem [{\citenamefont {D\'ora}\ and\ \citenamefont {Moca}(2025)}]{Dora2024}%
  \BibitemOpen
  \bibfield  {author} {\bibinfo {author} {\bibfnamefont {B.}~\bibnamefont
  {D\'ora}}\ and\ \bibinfo {author} {\bibfnamefont {C.}~\bibnamefont {Moca}},\
  }\bibfield  {title} {\bibinfo {title} {Momentum space nonstabilizerness for
  the transverse field quantum {Ising} model},\ }\href
  {https://doi.org/10.1103/mx8t-l4hf} {\bibfield  {journal} {\bibinfo
  {journal} {Phys. Rev. B}\ }\textbf {\bibinfo {volume} {112}},\ \bibinfo
  {pages} {125427} (\bibinfo {year} {2025})}\BibitemShut {NoStop}%
\bibitem [{\citenamefont {Ashida}\ and\ \citenamefont
  {Ueda}(2018)}]{Ashida2018}%
  \BibitemOpen
  \bibfield  {author} {\bibinfo {author} {\bibfnamefont {Y.}~\bibnamefont
  {Ashida}}\ and\ \bibinfo {author} {\bibfnamefont {M.}~\bibnamefont {Ueda}},\
  }\bibfield  {title} {\bibinfo {title} {Full-counting many-particle dynamics:
  Nonlocal and chiral propagation of correlations},\ }\href
  {https://doi.org/10.1103/PhysRevLett.120.185301} {\bibfield  {journal}
  {\bibinfo  {journal} {Phys. Rev. Lett.}\ }\textbf {\bibinfo {volume} {120}},\
  \bibinfo {pages} {185301} (\bibinfo {year} {2018})}\BibitemShut {NoStop}%
\bibitem [{\citenamefont {Verstraete}\ \emph {et~al.}(2009)\citenamefont
  {Verstraete}, \citenamefont {Cirac},\ and\ \citenamefont
  {Latorre}}]{verstraete2009}%
  \BibitemOpen
  \bibfield  {author} {\bibinfo {author} {\bibfnamefont {F.}~\bibnamefont
  {Verstraete}}, \bibinfo {author} {\bibfnamefont {J.~I.}\ \bibnamefont
  {Cirac}},\ and\ \bibinfo {author} {\bibfnamefont {J.~I.}\ \bibnamefont
  {Latorre}},\ }\bibfield  {title} {\bibinfo {title} {Quantum circuits for
  strongly correlated quantum systems},\ }\href
  {https://doi.org/10.1103/PhysRevA.79.032316} {\bibfield  {journal} {\bibinfo
  {journal} {Phys. Rev. A}\ }\textbf {\bibinfo {volume} {79}},\ \bibinfo
  {pages} {032316} (\bibinfo {year} {2009})}\BibitemShut {NoStop}%
\bibitem [{\citenamefont {Cervera-Lierta}(2018)}]{Cervera}%
  \BibitemOpen
  \bibfield  {author} {\bibinfo {author} {\bibfnamefont {A.}~\bibnamefont
  {Cervera-Lierta}},\ }\bibfield  {title} {\bibinfo {title} {Exact {I}sing
  model simulation on a quantum computer},\ }\href
  {https://doi.org/10.22331/q-2018-12-21-114} {\bibfield  {journal} {\bibinfo
  {journal} {{Quantum}}\ }\textbf {\bibinfo {volume} {2}},\ \bibinfo {pages}
  {114} (\bibinfo {year} {2018})}\BibitemShut {NoStop}%
\bibitem [{\citenamefont {Kattel}\ \emph {et~al.}(2023)\citenamefont {Kattel},
  \citenamefont {Pasnoori},\ and\ \citenamefont {Andrei}}]{Kattel2023}%
  \BibitemOpen
  \bibfield  {author} {\bibinfo {author} {\bibfnamefont {P.}~\bibnamefont
  {Kattel}}, \bibinfo {author} {\bibfnamefont {P.~R.}\ \bibnamefont
  {Pasnoori}},\ and\ \bibinfo {author} {\bibfnamefont {N.}~\bibnamefont
  {Andrei}},\ }\bibfield  {title} {\bibinfo {title} {Exact solution of a
  non-{Hermitian} {$\mathcal{PT}$}-symmetric spin chain},\ }\href
  {https://doi.org/10.1088/1751-8121/ace56e} {\bibfield  {journal} {\bibinfo
  {journal} {J. Phys. A: Math. Theor.}\ }\textbf {\bibinfo {volume} {56}},\
  \bibinfo {pages} {325001} (\bibinfo {year} {2023})}\BibitemShut {NoStop}%
\bibitem [{\citenamefont {Turkeshi}\ \emph
  {et~al.}(2025{\natexlab{b}})\citenamefont {Turkeshi}, \citenamefont
  {Tirrito},\ and\ \citenamefont {Sierant}}]{Turkeshi2025}%
  \BibitemOpen
  \bibfield  {author} {\bibinfo {author} {\bibfnamefont {X.}~\bibnamefont
  {Turkeshi}}, \bibinfo {author} {\bibfnamefont {E.}~\bibnamefont {Tirrito}},\
  and\ \bibinfo {author} {\bibfnamefont {P.}~\bibnamefont {Sierant}},\
  }\bibfield  {title} {\bibinfo {title} {Magic spreading in random quantum
  circuits},\ }\bibfield  {journal} {\bibinfo  {journal} {Nat. Commun.}\
  }\textbf {\bibinfo {volume} {16}},\ \href
  {https://doi.org/10.1038/s41467-025-57704-x} {10.1038/s41467-025-57704-x}
  (\bibinfo {year} {2025}{\natexlab{b}})\BibitemShut {NoStop}%
\bibitem [{\citenamefont {Tirrito}\ \emph
  {et~al.}(2025{\natexlab{b}})\citenamefont {Tirrito}, \citenamefont
  {Tarabunga}, \citenamefont {Bhakuni}, \citenamefont {Dalmonte}, \citenamefont
  {Sierant},\ and\ \citenamefont {Turkeshi}}]{Tirrito2025a}%
  \BibitemOpen
  \bibfield  {author} {\bibinfo {author} {\bibfnamefont {E.}~\bibnamefont
  {Tirrito}}, \bibinfo {author} {\bibfnamefont {P.~S.}\ \bibnamefont
  {Tarabunga}}, \bibinfo {author} {\bibfnamefont {D.~S.}\ \bibnamefont
  {Bhakuni}}, \bibinfo {author} {\bibfnamefont {M.}~\bibnamefont {Dalmonte}},
  \bibinfo {author} {\bibfnamefont {P.}~\bibnamefont {Sierant}},\ and\ \bibinfo
  {author} {\bibfnamefont {X.}~\bibnamefont {Turkeshi}},\ }\href
  {https://arxiv.org/abs/2506.12133} {\bibinfo {title} {Universal spreading of
  nonstabilizerness and quantum transport}} (\bibinfo {year}
  {2025}{\natexlab{b}}),\ \Eprint {https://arxiv.org/abs/2506.12133}
  {arXiv:2506.12133 [quant-ph]} \BibitemShut {NoStop}%
\end{thebibliography}%

\end{document}